# Novel Concept of Non-Debye Dipole Relaxation Processes for the Interpretation of Physical Origin of Dielectric Loss in the Glass Formers, Drugs, Polymers and Plastic Crystals

## G. Govindaraj

*Department of Physics, School of Physical, Chemical and Applied Sciences, Pondicherry University, R. V. Nagar, Kalapet, Puducherry 605 014, India,*
*e-mail: ggdipole@gmail.com*

The physical origin of dielectric loss is shown to be sum of n number of subunits relaxation of a molecule, where n=1,2,3.. For each subunit relaxation, the idea of intermolecular dipole-dipole interactions triggered non-Debye dipole, $(\mathbf{G})_n=((1-g_d)\mathbf{G}_0)_n$, and the ensuing dual dipole $(\mathbf{G}_\pm)_n=(\mathbf{G}_0\pm\mathbf{G})_n$, relaxation processes is proposed, where $\mathbf{G}_-=g_d\mathbf{G}_0$, $\mathbf{G}_+=(2-g_d)\mathbf{G}_0$, and $\mathbf{G}_0$ is a Debye dipole. Each subunit motion is statistically highly independent process and discriminated by Debye and non-Debye relaxation (NDR) time, where $g_d$ is an exponent $0<g_d<1$ and signifies interaction strength with a redistribution and conservation of Debye dielectric loss energy. The proposed concept provides a new insight for the NDR and discloses the physical origin of α, β, γ, δ relaxations and excess wing of glass formers, plastic crystals, drugs, *etc*., with an excellent agreement with experimental results.

In general, the dielectric loss of glass formers, supercooled liquids and solids do not exhibit the Debye responses except in some liquid dielectrics. The deviation from Debye response is referred as non-Debye relaxation (NDR) or many-body relaxation. Over the past several decades, many empirical relaxation laws or relationships have been proposed for NDR [1,2]. Among them, the time-honored models are: (i) Cole-Cole (CC) function $\epsilon_{CC}^*\propto[(1+i\omega\tau_{CC})^{1-\alpha_{CC}}]^{-1}$, $0\le\alpha_{CC}<1$ [3], (ii) the Cole-Davidson (CD) function $\epsilon_{CD}^*\propto[(1+i\omega\tau_{CD})^{\beta_{CD}}]^{-1}$, $0<\beta_{CD}\le1$, [4] (iii) Havriliak-Negami (HN) function $\epsilon_{HN}^*\propto[(1+(i\omega\tau_{HN})^{1-\alpha_{CC}})^{\beta_{CD}}]^{-1}$, $0<\beta_{CD}\le1$ [5], (iv) Kohlrausch-Williams-Watts (KWW) stretched exponential function $\phi(t)=\exp[-(t/\tau)^{\beta_{KWW}}]$, $0<\beta_{KWW}\le1$ [6,7], (v) Jonscher's "universal dielectric response (UDR)" [7] and (vi) Ngai's coupling model (CM) [1,2,7].

Dielectric loss spectra of glass forming materials are dominated by the structural α relaxation peak, which is significantly slowed down when the glass transition temperature $T_g$ is approached. In many glass-forming materials, besides the α relaxation peak, there is clear evidence of the presence of a higher frequency relaxation, commonly called as β, γ and δ relaxations, because they give rise to an additional peaks in dielectric loss [1]. In general, the combinations of functions (i)-(vi) listed above are used to describe the α, β, γ and δ relaxations. Johari and Goldstein [8] proposed that β relaxation (or secondary relaxation) may be a universal and intrinsic feature of glass forming liquids at low temperatures, since it is observed even for several rigid molecules. This universality is, however, apparently inconsistent and diluted by the following facts [9]. There are several glass formers that do not display a well-resolved β relaxation and usually shows the excess wing (EW) on the high-frequency side of the α relaxation, which is described as a second power law $\epsilon''(\omega) \approx \omega^{-b}$, where the exponent b is lying $0 < b < \beta_{KWW}$ [10,11].

Until now, the physical origin of β, γ and δ relaxations and the EW is a puzzle and remains unclear, and several controversial results are reported [10-16] on the glass formers classifying as 'type-A' without β process but showing an EW, and 'type-B' with β . Therefore, β relaxation and EW is still unknown fundamental process in interacting systems. Capturing this fundamental process is a worthwhile scientific attempt to solve glassy dynamics, in particular, and NDR in general. In the present letter, for the first time, the concept of intermolecular dipole-dipole interaction triggered non-Debye dipole and an ensuing dual dipole is introduced to study the α, β, γ, δ relaxations and EW in terms of subunits relaxation process of a molecule. Dielectric and relaxation functions are obtained for the dual dipole process. A meticulous study of dielectric loss spectra of 12 different systems has been performed and the results on the prototypical glass former glycerol and 2-ethyl-1-hexanol are reported.

Let the condensed matter system consist of reorientation of dipolar entities with the dipole moment $(\mathbf{G_0})_n$, where n is number of subunits of a molecule with dipole moments $(\mathbf{G_0})_1$, $(\mathbf{G_0})_2$, $(\mathbf{G_0})_3$, …. $(\mathbf{G_0})_n$, and n depends of nature of the molecule and number of n contributing to loss depends on temperature. The bold face in the text indicates vector quantity. Let us consider the subunit 1 of a molecule with dipole moment $(\mathbf{G_0})_1$ (for example three subunits in ethanol -$CH_3$-, -$CH_2$-,-OH-). In an electric field, under a given thermodynamic condition, instantaneous transfer of dipole moment $\mathbf{G_0}$ (suffix 1 is dropped for clarity) is not possible due to intermolecular interactions, and hence a dipole $\pm\mathbf{G}=(1-g_d)\mathbf{G_0}$ is triggered in terms of $\mathbf{G_0}$, where $0 < g_d \leq 1$, $\mathbf{G_0}$ ($g_d=1$) and $\pm\mathbf{G}$ are called as a Debye and a non-Debye dipoles. The effect of $\pm\mathbf{G}$ on $\mathbf{G_0}$ is the creation of a dual dipole:
$$\mathbf{G_+}=\mathbf{G_0}+\mathbf{G}=(2-g_d)\mathbf{G_0}, \text{ and } \mathbf{G_-}=\mathbf{G_0}-\mathbf{G}=g_d\mathbf{G_0}, \tag{1}$$
where the number density of $\mathbf{G_0}$ is N and $\mathbf{G_\pm}$ is N/2. The dipole moments of $\mathbf{G_+}$ and $\mathbf{G_-}$ is increased and decreased by a factor of $(1-g_d)$ with respect to $\mathbf{G_0}$ [17]. In the glass former, as T approaches glass transition temperature $T_g$, the glass former of a molecule with n subunits and their interaction ensued dual dipoles generate energy landscaping, *i.e*, a complicated dependence of energy on configuration, a change in configurational entropy and a progressive increase of viscosity related to fragility.

Although many-particle effect is undoubtedly play an important role and it is interesting and instructive to clarify to what extend one can describe the essential aspects of the NDR in the framework of single-particle process. In the proposed model, the $(\mathbf{G_0})_n$ is treated as 'single-particle like', and the dual dipole $(\mathbf{G_\pm})_n$ is treated as 'many-particle like' involving all the molecules and their n subunits with interaction strength $0 < (g_d)_n < 1$. Therefore, the $(\mathbf{G_\pm})_n$ relaxation dynamics with interaction strength $0 < (g_d)_n < 1$ defines a new NDR.

Langevin function is obtained for the $\mathbf{G_\pm}$ and it is found to be:
$$<\mu_i>=\mu_i L(z_i), \, L(z_i) = coth(z_i) - 1/z_i, \tag{2}$$
where $z_i = \mu_i E/(k_B T)$, E is external applied electric field, the symbols $<>$ stand for ensemble average for the dipoles and $\mu_i$ stands for the dipoles $\mathbf{G_0}$, $\mathbf{G_+}$ and $\mathbf{G_-}$. The statistical distribution of irreversible processes follows the Boltzmann factor $exp[-(U_i/k_BT)]$, $U_i=G_icos(\theta_i)E$, where the energy of $\mathbf{G_0}$, is redistributed through, $\mathbf{G_-}$, and $\mathbf{G_+}$, such that total energy is conserved, and $\theta_i$ is angle between dipole moment and **E**. For $z_i>>1$ the $L(z_i)=1-1/z_i$ and approaches to one, however, for low field limit $z_i<<1$, the linear regime $L(z_i)=(1/3)z_i$ and saturation *depends* on $g_d$. These features are shown as inset in Fig. 1(a).

The Debye dielectric function $[\epsilon_d^*(\omega)]_{G_0}$ is given by:

$$[\epsilon_d^*(\omega)]_{G_0} - \epsilon_\infty = (\Delta\varepsilon)_{G_0}/(1 + i\omega\tau_D) = (\Delta\varepsilon/(1 + s\Gamma))_{G_0}, \quad s = i\omega, \ (\Gamma)_{G_0} = \tau_D, \tag{3}$$

where $(\Delta\epsilon)_{G_0} = (\epsilon_s - \epsilon_\infty)_{G_0} = \dfrac{NG_0^2}{3\epsilon_0 k_B T} - \epsilon_\infty$, $\epsilon_s$ and $\epsilon_\infty$ are the high and low external frequency dielectric limits respectively. The effect of $\mathbf{G}$ on $[\epsilon_d^*(\omega)]_{G_0}$, is shifting of the Debye term, $i\omega\tau_d$ by a factor of $1/(i\omega\tau_d)^{1-g_d}$, and hence, $i\omega\tau_d$ becomes $i\omega\tau_d/(i\omega\tau_d)^{1-g_d} = (i\omega\tau_d)^{g_d}$, where $\tau_d$ is assigned as relaxation time when $0 < g_d < 1$, and as $\tau_D$ for $g_d = 1$. The dielectric function $[\epsilon_d^*(\omega)]_{G_-}$ is obtained as:

$$[\epsilon_d^*(\omega)]_{G_-} - \epsilon_\infty = \frac{\Delta\epsilon}{1 + (i\omega\tau_d)^{g_d}} = \left(\frac{\Delta\epsilon}{1 + s\Gamma}\right)_{G_-}, \quad (S)_{G_-} = \frac{s}{s^{1-g_d}}, \ (\Gamma)_{G_-} = \tau_d^{g_d}. \tag{4}$$

This is similar to Cole-Cole type dielectric function. However, the dielectric strength for $\mathbf{G}_-$ is found to be $(\Delta\epsilon)_{G_-} = (\epsilon_s - \epsilon_\infty)_{G_-} = \left(\dfrac{N}{2}\dfrac{G_0^2}{3\epsilon_0 k_B T}\right)_{G_-} - \epsilon_\infty$, N/2 is the density of $\mathbf{G}_-$ dipoles. Interesting observation is, when $0 < g_d < 1$, $(\Delta\epsilon)_{G_-}$ is smaller than $(\Delta\epsilon)_{G_0}$, by factor of dual dipole dielectric loss (½) $\tan(g_d\pi/4)$ at $\omega = 1/\tau_d$.

The dielectric function $[\epsilon_d^*(\omega)]_{G_+}$ is obtained by incorporating the effect of $\mathbf{G} = -(1-g_d)\mathbf{G}_0$ on $\mathbf{G}_0$. The influence of $\mathbf{G}$ on $[\epsilon_d^*(\omega)]_{G_0}$, $(g_d = 1)$ is the shifting of the Debye terms, $i\omega\tau_d$ by a factor of $1/((-i)\omega\tau_d)^{-(1-g_d)}$ and hence, $i\omega\tau_d$ becomes $i\omega\tau_d/((-i)\omega\tau_d)^{-(1-g_d)} = i^{g_d}(\omega\tau_d)^{2-g_d}$, since $(-i)^{-(1-g_d)} = i^{(1-g_d)}$. The $[\epsilon_d^*(\omega)]_{G_+}$ is obtained as:

$$[\epsilon_d^*(\omega)]_{G_+} = \frac{\Delta\epsilon}{1 + i^{g_d}(\omega\tau_d)^{2-g_d}} = \left(\frac{\Delta\epsilon}{1 + s\Gamma}\right)_{G_+}, \quad (S)_{G_+} = \frac{s}{(-s)^{-(1-g_d)}}, \ (\Gamma)_{G_+} = \tau_d^{2-g_d}, \tag{5}$$

where $(\Delta\epsilon)_{G_+} = (\epsilon_s - \epsilon_\infty)_{G_+} = \left(\dfrac{N}{2}\dfrac{G_0^2}{3\epsilon_0 k_B T}\right)_{G_+} - \epsilon_\infty$. Since, the dielectric loss energy of $\mathbf{G}_0$ is redistributed in equal magnitudes on lower and higher sides through $\mathbf{G}_\pm$, the dielectric loss in $[\epsilon_d^*(\omega)]_{G_\pm}$, spreads with respect to $[\epsilon_d^*(\omega)]_{G_0}$. Here, again, as indicated in $(\Delta\epsilon)_{G_-}$, when $0 < g_d < 1$, $(\Delta\epsilon)_{G_+}$ is smaller than $(\Delta\epsilon)_{G_0}$.

According to proposed model, for each subgroup motion, $\tau_D = \tau_d$ ever, when $0 < g_d < 1$, then, what is NDR time? Now, the dielectric loss of $\mathbf{G}_0$ is maximum at $\tau_D = 1/\omega_D$ $(g_d = 1)$, where $\omega_D$ is the Debye loss peak frequency. Equating the dielectric loss energy of $\mathbf{G}_0$ with dielectric loss of $\mathbf{G}_-$ or $\mathbf{G}_+$ at $\omega = 1/\tau_d$, yields slow and fast NDR times with respect to $\tau_D$ and these are:

$$\tau_s^* = 1/\omega_s^* = \tau_D(c + \sqrt{c^2 - 1}), \tag{6}$$
$$\tau_f^* = 1/\omega_f^* = \tau_D(c - \sqrt{c^2 - 1}), \tag{7}$$

where $c = \cot(g_d\pi/4)$, dielectric loss at the loss peak. The $\tau_s^*$ and $\tau_f^*$ are roots of symmetric Debye loss curve with respect to the loss peak $\tau_D$, and with these roots, the real part of $\epsilon_{GG}^*(\omega)$ and $\phi_{GG}(t)$ show *hysteresis* like structure as a function of $g_d$ and, no hysteresis and Debye result for $g_d = 1$.

The relaxation functions for the $\mathbf{G}_0$, $\mathbf{G}_-$ and $\mathbf{G}_+$ dipole processes are:

$$[\phi_d(t)]_{G_0} \propto exp[-(t/\tau_D)], \propto exp\left[-(T/\Gamma)_{G_0}\right], \ (T)_{G_0} = t \tag{8}$$
$$[\phi_d(t)]_{G_-} \propto exp[-(t/\tau_d)^{g_d}] \propto exp\left[-(T/\Gamma)_{G_-}\right], \ (T)_{G_-} = t^{g_d} \tag{9}$$
$$[\phi_d(t)]_{G_+} \propto exp[-(t/\tau_d)^{2-g_d}] \propto exp\left[-(T/\Gamma)_{G_+}\right], \ (T)_{G_+} = t^{2-g_d} \tag{10}$$

where Eqs. (3)-(5) and (8)-(10) are Laplace or Fourier transform pairs in T and S domains, where both $t/\tau_d$ and $\omega\tau_d$ are compressed and stretched by $g_d$ in an equal magnitudes. The relaxation and dielectric functions with slow and fast NDR time become;

$$[\phi_d(t)]_{G_-} \propto \begin{cases} exp[-(t/\tau_s^*)^{g_d}] \\ exp\left[-(t/\tau_f^*)^{g_d}\right] \end{cases} ; \ [\phi_d(t)]_{G_+} \propto \begin{cases} exp[-(t/\tau_s^*)^{2-g_d}] \\ exp\left[-(t/\tau_f^*)^{2-g_d}\right] \end{cases},$$

$$[\epsilon_d^*(\omega)]_{G_-} \propto \begin{cases} 1/(1+(i\omega\tau_s^*)^{g_d}) \\ 1/(1+(i\omega\tau_f^*)^{g_d}) \end{cases}; [\epsilon_d^*(\omega)]_{G_+} \propto \begin{cases} 1/(1+i^{g_d}(\omega\tau_s^*)^{2-g_d}) \\ 1/(1+i^{g_d}(\omega\tau_f^*)^{2-g_d}) \end{cases}. \tag{11}$$

When n subunits of a molecule contributes to the dielectric spectra, then the dielectric and relaxation function become:

$$\epsilon_{GG}^*(\omega) = \sum_{m=1}^{n}([\epsilon_d^*(\omega)]_{G_-} + [\epsilon_d^*(\omega)]_{G_+})_m, \tag{12}$$

$$\phi_{GG}(\omega) \propto \sum_{m=1}^{n}([\phi_d(t)]_{G_-} + [\phi_d(t)]_{G_+})_m. \tag{13}$$

In the presence of free charge conduction, $\sigma_{dc}/i\omega\epsilon_0$, is added. For each subunit motion, $(\tau_D)_m$ is the primary relaxation time of $(G_0)_m$, $(\tau_s^*)_m$ and $(\tau_f^*)_m$ are NDR time of $(G_-)_m$ or $(G_+)_m$ or $((G_-)+(G_+))_m$ and this is the *novel result* on the relaxation dynamics of the proposed model.

The relaxation time of glass formers shows a deviation from Arrhenius law and it is parameterized with Vogel-Fulcher-Tamman (VFT) equation [18] for each subunit motion as:

$$(\tau_d)_m = (\tau_0 exp(A_0 T_0/(T-T_0)))_m, \text{ for } T > T_g, \tag{14}$$

where m=1, 2, …, n, $T_0$ is the VFT approximation of the ideal glass transition temperature, $A_0$ is the strength parameter, $\tau_0$ is a pre-factor of the order inverse phonon frequency and further characterized based on fragility index, $m_p = log_{10}(e)(A_0(T_0/T_g)(1-T_0/T_g)^{-2})_m$ [18], where Arrhenius equation is $(\tau_d)_m = (\tau_0 exp(E_a/k_B T))_m$ with $E_a$ as activation energy.

Dielectric spectra of glycerol (Gly, $C_3H_5(OH)_3$, $CH_2$-OH-CHOH-$CH_2$OH, $T_g$=193K) and 2-ethyl-1-hexanol (2E1H, $C_8H_{17}OH$, $C_2H_5$-$(CH_2)_2$ -CH-$CH_2$OH-$C_2H_5$, $T_g$=147K) are analyzed based on proposed model. As measured data are collected from 'Glass and Time: Data Repository' [19-21]. Dielectric loss, $\epsilon_{GG}''(\omega)$, and constant, $\epsilon_{GG}'(\omega)$, are fitted simultaneously in the temperature range 192 to 252K in steps of 4K for Gly and 157 to 226K in steps of 1.5K (5K in higher T) for 2E1H using Eq. (12). From low to high temperatures, the number of subunits contributed to the $\epsilon_{GG}''(\omega)$ for Gly are: 192-224K, n=4; 228K, n=3; 232-244K n=2; 248-252K n=1, and for 2E1H: 157-165.5, n=4; 167-175K, n=3; 175-224.5K n=2; 226K, n=1; respectively. Good quality fits are shown in Figs. 1 (a) & (b) for the Gly and 2(a) & (b) for the 2E1H, as color lines and black dots are data. In both systems, no loss contribution from $(G_+)_m$, since in Langevin function, Eq. (2), in low temperature region, the linear regime is dominant by $(G_-)_m$, as shown in inset Fig.1(a). However, there are systems, $(G_+)$ dielectric loss contribution is observed and some of them are shown in Table 1. The $\epsilon_{GG}''(\omega)$ and $\epsilon_{GG}'(\omega)$ are shown for each subunits motion and their sum in Fig. 3(a) & (b) for Gly and 2E1H, for T=161.5 and 196K. The temperature dependence $(\tau_d)_n$ (n=1,2,3,4) are shown in Figs. 4(a) & (b).

The red, blue, magenta and purple arrows on the curve in Figs. (1)-(3) indicate $(\tau_d)_n$ (n=1,2,3,4) in decreasing order. The inset table in Fig. 3(b) shows fit parameters (FP) estimate and their standard error (SE) and the inset tables in Fig. 3(a) shows how many times the $(\tau_d)_1$ higher than fast subgroup motions in Gly and 2E1H. The schematic molecular structure of Gly and 2E1H structure are shown in Figs. 1, 2 & 3. In the sketch of the molecular structures, the oxygen atom is highlighted in red.

It is clear from inset tables in Figs. 3(a)-(b), the shape of $\epsilon_{GG}''(\omega)$ curves depends on the closeness of subunits relaxation time $(\tau_d)_1$-$(\tau_d)_4$, magnitude of $(\Delta\epsilon)_1$-$(\Delta\epsilon)_4$ of $(G_0)_{1,2,3,4}$ and interaction strength $(g_d)_1$-$(g_d)_4$. In the case of Gly, for n=4, in T=192-224K range, relaxation times $(\tau_d)_1$-$(\tau_d)_4$ are closely spaced $((\tau_d)_1/(\tau_d)_4\sim250)$, whereas, in the case of 2E1H, for n=4, in T=157-165.5K range, $(\tau_d)_1$-$(\tau_d)_4$ are well separated $((\tau_d)_1/(\tau_d)_4\sim10^6)$ as indicated in the inset table in Fig. 3(a). The insets in Fig. 4(a) &(b) show the strength of interaction $(g_d)_1$-$(g_d)_4$, the dielectric strength $(\Delta\epsilon)_1$-$(\Delta\epsilon)_4$ of $(G_0)_{1,2,3,4}$ and their sum, and $\epsilon_\infty$ for Gly and 2E1H. The insets in Fig. 4(a) & (b) show that for both Gly and 2E1H, $(\tau_d)_1$ is the slowest and Debye relaxation time $(\tau_D)_1$, since $(g_d)_1$=1. This is the first report showing 4 different closely spaced relaxation times in Gly and slowest motion as a Debye relaxation, similar to 2E1H [22,23] and other several classes of hydrogen-bonded liquids [24]. Then, as per the existing nomenclatures, $(\tau_d)_1$-$(\tau_d)_4$, are Debye, $\alpha$, $\beta$, and $\gamma$ relaxations respectively. Therefore, the structural $\alpha$ relaxation is $(\tau_d)_2$, mysterious JG's $\beta$ relaxation and $\gamma$ relaxations are motion of subunits of a molecule having relaxation times $(\tau_d)_3$ and $(\tau_d)_4$.

The inset tables and Figs. 4(a) & (b) showing VFT fit results and fragility parameters for $(\tau_d)_n$ (n=1, 2, 3, 4) for Gly, $(\tau_d)_{1,2}$ and $(\tau_d)_{3,4}$ showing VFT and Arrhenius fit respectively for 2E1H. Now, what is EW? Is EW a different phenomenon altogether? Is EW high frequency flank of loss peak caused by $\beta$ or $\gamma$? According to the model, whenever, subunits' $(\tau_d)_n$ (n=1, 2, 3,.. or $\alpha$, $\beta$, $\gamma$, .. ) are well separated and if the interaction strength is weak $((g_d)_n \gtrsim 0.6)$, the loss peak is observed and however, there is no EW, as in 2E1H in Fig. 3(a). On the other hand, $(\tau_d)_n$ are closely spaced and if the interaction strength is strong $((g_d)_n \lesssim 0.6)$, loss peaks are merged and the EW is observed, as in Gly in Fig. 3(a). The extent of EW depends on interaction strength, for example, in 2E1H, $(g_d)_4 = 0.914$ for $\gamma$ $(\tau_d)_4$ process, whereas, in Gly, $(g_d)_4 = 0.256$ for $\gamma$ $(\tau_d)_4$ process. Therefore, EW is not a different phenomenon, and it is high frequency flank of loss peak caused by $\alpha$ or $\beta$ or $\gamma$. Further, dielectric loss [25] data of ten systems are analyzed and number of subgroups contributed to the loss are shown in Table 1.

In summary, the NDR in dielectric spectra of glass former is modelled by considering a molecule having n number of subunits and their relaxations. For each subunit, intermolecular interaction triggered non-Debye dipole and ensuing dual dipole relaxation is predicted in terms of Debye relaxation time $\tau_D$, and its NDR time $\tau_s^*$ and $\tau_f^*$, considering redistribution and conservation Debye dielectric loss energy. According to the proposed model, EW is a interaction strength $(0 < g_d < 1)$ dependent high flank of $\alpha$, $\beta$, $\gamma$, $\delta$ and it is not a different phenomenon and hence the classification of type-A and type-B glass formers is superfluous. The relaxation time $(\tau_d)_m(T \rightarrow T_g)$ and $(g_d)_m(T \rightarrow T_g)$, m=1,2,3,...,n indicates the strong interaction initiated dynamic heterogeneity [27] $i.e.$, spatially varying time scales of molecular rearrangements in a single component system, having extremely slow relaxations, and when $T >> T_g$, n becomes 1 or 2. The proposed concept encompasses and amends the models of CC, KWW, UDR, CM and have the features of CD and HN when $(\tau_d)_m$ are closely spaced. It remains to be seen the subunits type (example $CH_2$-OH–CH-OH–$CH_2$-OH) and its $(\tau_d)_m$ and these requires further experiments. Presence of dual dipole dynamics in ac conductivity and shear mechanical relaxation has been realized and these are our future publications.

## ACKNOWLEDGMENTS

Author thanks, the group leader, Prof. Jeppe Dyre, for making availability of dielectric data "Glass and Time: Data Repository" online at http://glass.ruc.dk/data. Author acknowledges grant in aid of CSIR project F.03(1279)/13/EMR-II.

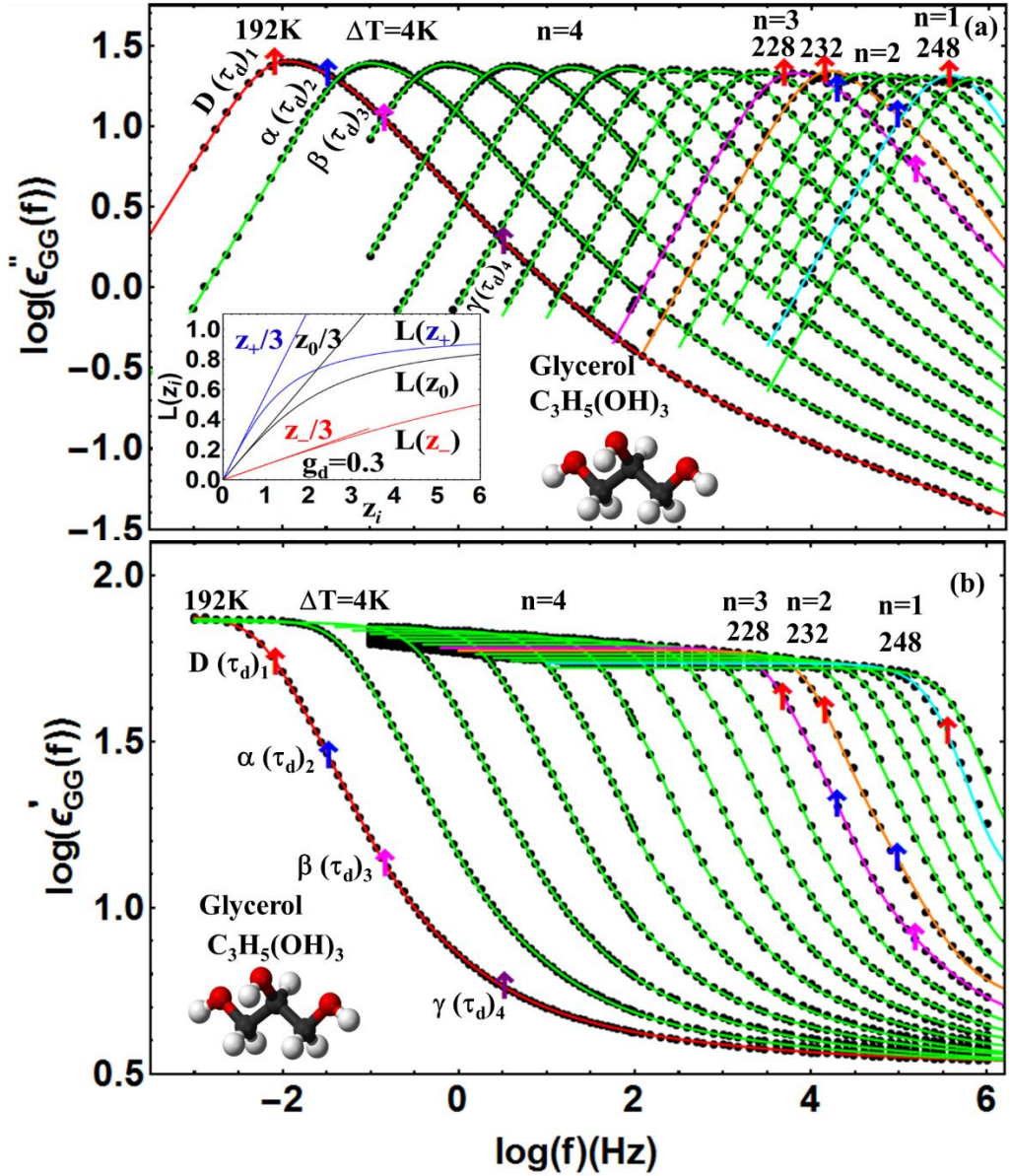

Fig. 1: (a) The color lines are fit results of dielectric loss and (b) dielectric constant of glycerol as a function of frequency obtained based on Eq. (12) for n=4 (red onwards), 3 (magenta), 2 (orange onwards), 1 (cyan onwards) for T=192-252K, ΔT=4K. Arrows indicate relaxation time: n=1, Debye, $(\tau_d)_1$; n=2, α, $(\tau_d)_2$; n=3, β,$(\tau_d)_3$; n=4, γ, $(\tau_d)_4$ of each subunit motion. Inset plot in Fig.1(a) shows Langevin Eq. (2) as function of $z_i=\mu_i E/k_B T$, $\mu_i$=G$_0$, G$_-$, and G$_+$ for $g_d$=0.3. For further details refer the text.

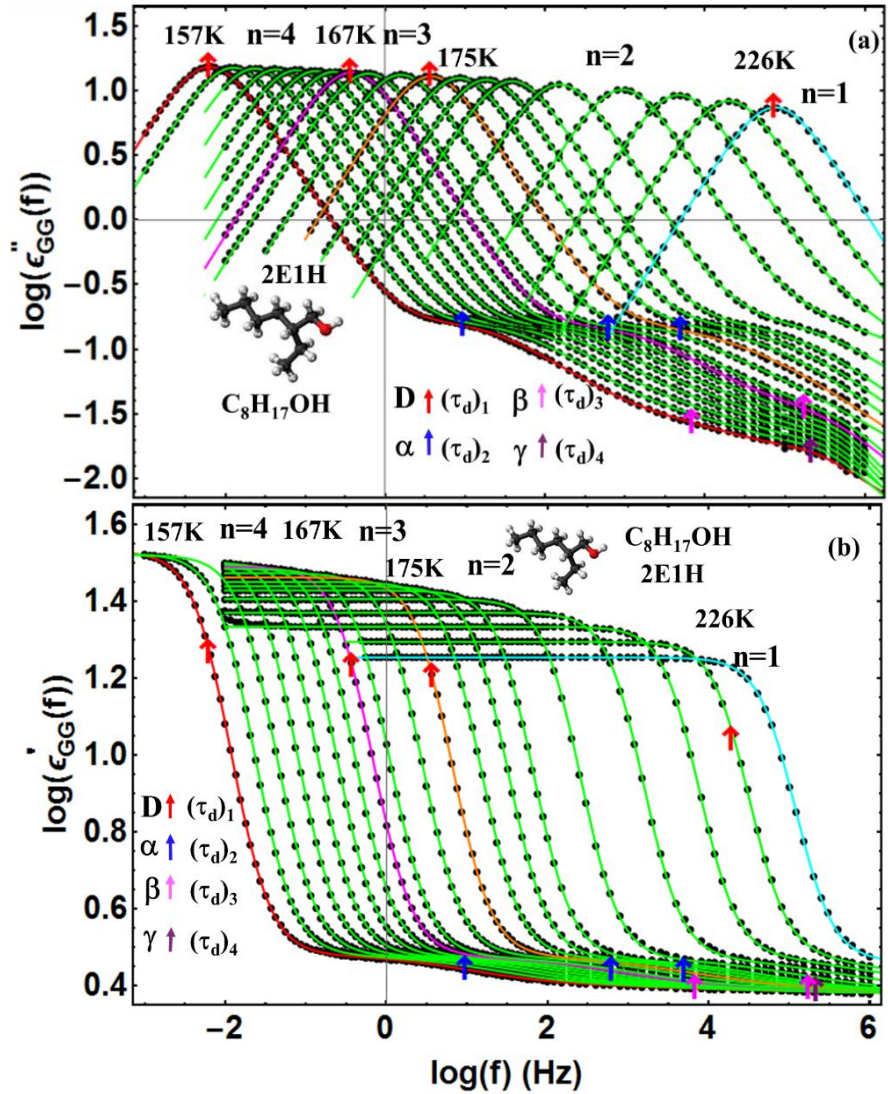

Fig. 2: (a) The color lines are fit results of dielectric loss and (b) dielectric constant of 2E1H as a function of frequency based on Eq. (12) for n=4 (red onwards), 3 (magenta onwards), 2 (orange onwards), 1 (cyan) for T=157-226K, ΔT=1.5K. Arrows indicate relaxation time: n=1, Debye, $(\tau_d)_1$; n=2, α, $(\tau_d)_2$; n=3, β, $(\tau_d)_3$; n=4, γ, $(\tau_d)_4$; of each subunit motion.

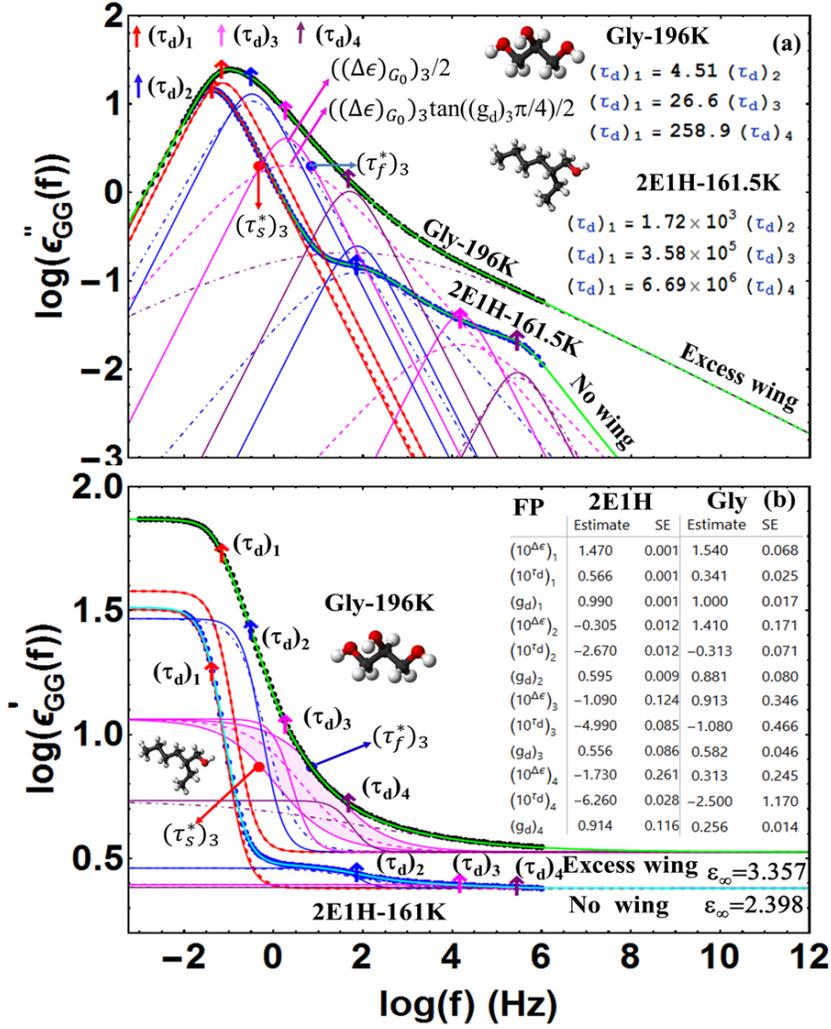

Fig. 3: (a) Fit results for the dielectric loss and (b) dielectric constant for the Gly and 2E1H as function of frequency based on Eq. (12) for n=4, where n=1, Debye, $(\tau_d)_1$; n=2, α, $(\tau_d)_2$; n=3, β, $(\tau_d)_3$; n=4, γ, $(\tau_d)_4$. In (a), both $(\mathbf{G_0})_{1,2,3,4}$ and $(\mathbf{G_-})_{1,2,3,4}$ contributions are shown as solid line and dashed, dot dashed lines. For Gly, in (a) dielectric strength of $(\mathbf{G_0})_3$ is $(\Delta\epsilon)_3 = (\Delta\epsilon)_{\mathbf{G_0}}/2$ and $(\mathbf{G_-})_3$ is $(\Delta\epsilon)_{\mathbf{G_-}} = (\Delta\epsilon)_3 \tan((g_d)_3\pi/4)$. For Gly, in (b), shaded magenta region is hysteresis obtained for $(\mathbf{G_-})_3$ with slow (red dot) & fast (blue dot) relaxation time based on Eqs. (11). Further details refer the text.

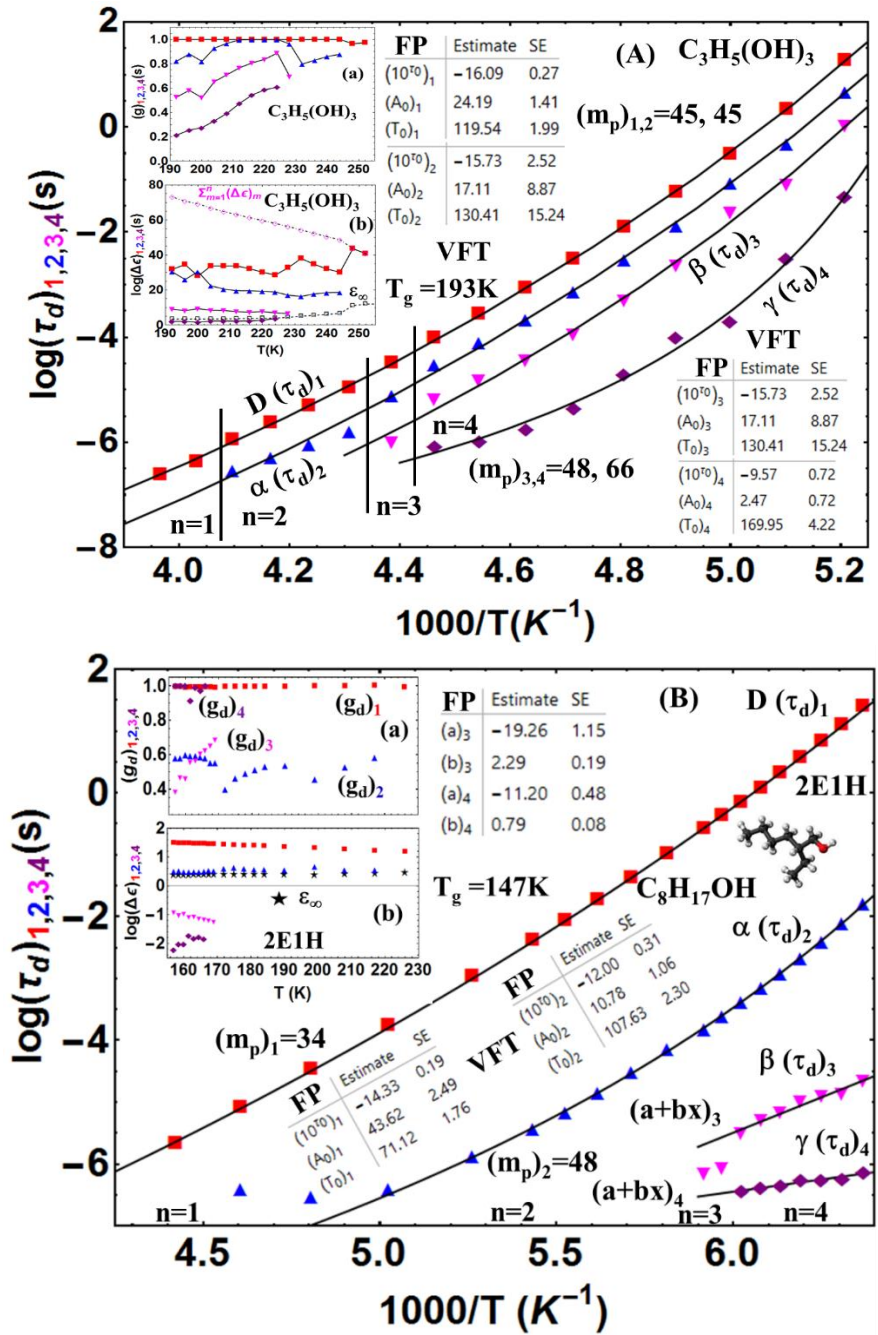

Fig. 4: Temperature dependence $(\tau_d)_1$-$(\tau_d)_4$ obtained based on Eq. (12) for (A) Gly and (B) 2E1H. The insets of (A) and (B) show temperature dependence $(g_d)_1$-$(g_d)_4$ (a) and Debye dielectric strength $(\Delta\varepsilon)_1$-$(\Delta\varepsilon)_4$ (b) (in log scale for 2E1H) of subunits, their sum and $\varepsilon_\infty$. The estimate of VFT parameters, fragility and estimate of Arrhenius fit parameters (FP) and standard error (SE) are indicated for $(\tau_d)_1$-$(\tau_d)_4$, where n=1, Debye; n=2, $\alpha$; n=3, $\beta$, n=4, $\gamma$.

Table 1: Different molecules and their subunits relaxation at different temperatures. g-glass, d-drug, p-polymer pc-plastic crystal, and *1*-CAN-*1*-cynoadamantane, PCNB-Pentachloronitrobenzene, AF-Arrhenius fit in the form, a+bx, a=$10^{\tau_0}$, b is slope and x=1000/T.

| S. No. | Molecule/type | Temp. range in (K) No. Temp. points | Existing relaxation nomenclature/ Number of subunits relaxation /dual dipole type | $T_g$ in K/ Fragility $(m_p)_n$/AF | Data Ref. |
|---|---|---|---|---|---|
| 1. | 2-butanol (g) | 128-191 (18) | D, α, β, δ/($\tau_d$)$_{1,2,3,4}$ / (**G**−)$_{1,2,3,4}$ | 120/38,51 -16.25+1.62x -10.62+0.57x | [19] |
| 2. | 2-Methylpentane 2,4 diol (g) | 265-278 (7) | α→D, β, δ/($\tau_d$)$_{1,2,3}$ /(**G**−)$_{1,2,3,4}$ | 187/49,59, 91 | [19] |
| 3. | 1-proponal (g) | 103.6-160.2 (15) | D, α, β/($\tau_d$)$_{1,2,3}$/(**G**−)$_{1,2,3}$ | 98/37, 86, −14.16+1.12x | [24] |
| 4. | Ethanol (g) | 96-231 (8) | α, β, δ/($\tau_d$)$_{1,2,3}$/(**G**−)$_{1,2,3}$ | 97/60, 89 | [16, 25] |
| 5. | Ketoprofen (d) | 273.15-343.15 (17) | D, α, β, δ/($\tau_d$)$_{1,2,3,4}$ / (**G**−)$_{1,2,3,4}$ | 270/ 77, 75, 84, 90 | [26] |
| 6. | Propylene carbonate (PC)(p) | 158-183 (6) | α→D, β, δ/($\tau_d$)$_{1,2,3}$ /(**G**−)$_{1,2,3}$ | 159/ 80,76, 93 | [11] |
| 7. | Ortho-carborane (*o*-CA) (pc) | 130-252 (7) | α, β /($\tau_d$)$_{1,2}$ / (**G**−)$_{1 \text{ or } 2}$ & (**G**±)$_{1 \text{ or } 2}$ | -18.77+2.77x -17.45+2.34x | [10] |
| 8. | Mata-carborane (*m*-CA) (pc) | 180-275 (6) | ($\tau_d$)$_{1,2}$/ α, β /(**G**−)$_{1,2}$ | -17.76+2.59x -18.73+2.60x | [10] |
| 9. | *1*-CAN (pc) | 260-420 (6) | α, β/($\tau_d$)$_{1,2}$/ (**G**−)$_{1,2}$ | -16.39+3.04x -16.95+2.61x | [10, 25] |
| 10. | PCNB (pc) | 190-415 (7) | α, β, δ/($\tau_d$)$_{1,2,3}$ /(**G**−)$_1$ & (**G**±)$_2$ | -15.99+3.54x -15.46+3.13x | [25] |

**Novel Concept of Non-Debye Dipole Relaxation Processes for the Interpretation of Physical Origin of Dielectric Loss in the Glass Formers, Drugs, Polymers and Plastic Crystals**

**Supplement materials**

**G. Govindaraj**

*Department of Physics, School of Physical, Chemical and Applied Sciences, Pondicherry University, R. V. Nagar, Kalapet, Puducherry 605 014, India*

In Figs. (1)-(6), the features of non-Debye dipole $(\mathbf{G})_n$, dual dipole $(\mathbf{G}_\pm)_n$, non-Debye relaxation and dielectric functions are explained by referring respective equation in the paper, where n is number of subunits of a molecule. The non-Debye dipole, and ensuing dual dipole, the dielectric and relaxation functions proposed in the present work is referred as GG model and dipole. For the ten different dielectric systems listed in Table 1 in the paper, the results are shown in Figs. (7)-(17). The data extracted from the analysis are shown in Table 1 in the paper. The temperature dependence data of dielectric strength and strength of interaction are joined in the figures to guide the eye. The best quality fit is demonstrated in Fig. 18 for the three systems by fitting the data with existing models and comparing with the present proposed model. In Fig. 18 (i) Fit results and fit residuals of HN+CC and present proposed GG (n=4) model are compared for glycerol (Gly) at 208K. (ii) Fit results and fit residuals of KWW+CC, CD+CC and present proposed GG (n=3) model are compared for propylene carbonate (PC) at 168K. (iii) Fit results and fit residues of KWW+CC, CD+CC and present proposed GG (n=2) model are compared for 1-cynaoadmaentane (*1*-CAN) at 260K, where y scale is shifted by 0.4 units for each model starting from GG. The least magnitude of fit residues is observed for the present proposed GG model. This result is consistent for different systems at different temperatures.

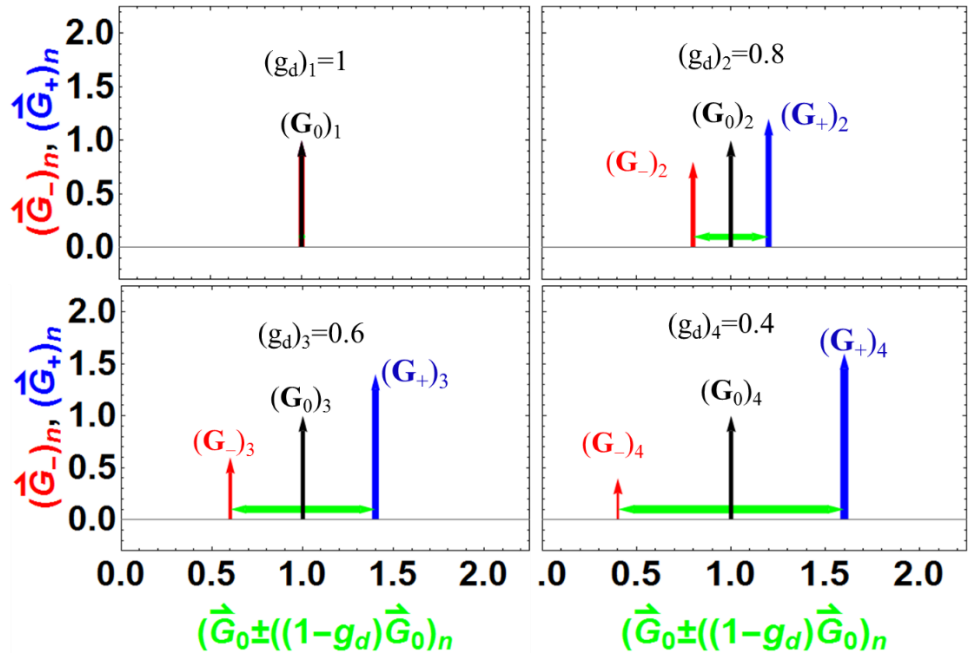

FIG. 1. Scheme showing formation of non-Debye dipole and ensuing dual dipole (Eq. (1)) for the subunits of a molecule with different strength of Debye dipole moments $(\mathbf{G_0})_1$, $(\mathbf{G_0})_2$, $(\mathbf{G_0})_3$, and $(\mathbf{G_0})_4$. The $g_d=1$ is a Debye process and $0<(g_d)_n<1$ is a non-Debye process with the intermolecular interaction strength $(g_d)_n$. The figures show $(\mathbf{G_\pm})_n=(\mathbf{G_0}\pm(1-g_d)\mathbf{G_0})_n$ for (i) $g_d=1$ (Debye), the non-Debye dipole is $(\mathbf{G})_1=0$, $(\mathbf{G_\pm})_1=(\mathbf{G_0})_1$ is a Debye, and there is no non-Debye dipole; (ii) $g_d=0.8$, the non-Debye dipole is $(\mathbf{G})_2=\pm(0.2\mathbf{G_0})_2$, dual dipole is $(\mathbf{G_-})_2=(0.8\mathbf{G_0})_2$, $(\mathbf{G_+})_2=(1.2\mathbf{G_0})_2$; (iii) $g_d=0.6$, the non-Debye dipole is $(\mathbf{G})_3=\pm(0.4\mathbf{G_0})_3$, dual dipole is $(\mathbf{G_-})_3=(0.6\mathbf{G_0})_3$, $(\mathbf{G_+})_3=(1.4\mathbf{G_0})_3$ and (iv) $g_d=0.4$, the non-Debye dipole is $(\mathbf{G})_4=\pm(0.6\mathbf{G_0})_4$, dual dipole is $(\mathbf{G_-})_4=(0.4\mathbf{G_0})_4$, $(\mathbf{G_+})_4=(1.6\mathbf{G_0})_4$. For example (a) in hydrogen bonded system $i.e.$, in ethanol, the possible subunits are (i) methyl (-CH$_3$-) (ii) methylene (-CH$_2$-) (iii) hydroxyl (OH), (b) in van der Waals bonded system $i.e.$, in the plastic crystals of carborane (chemical formula $C_2B_{10}H_{12}$, abbreviated as CA) have three isomeric state (i) in $para$-CA, the two carbon atoms are in opposite positions and hence no net dipole moment, (ii) in $meta$-CA, the two carbon atoms are in 1 and 7 positions, (iii) $ortho$-CA, the two carbon atoms 1 and 2 positions. In $ortho$ and $para$-CA have two subunits of different dipole moment strength due to different positions of carbon atoms, refer inset in Fig. 14 (a) for the molecular structure of $para$, $ortho$ and $meta$ CA.

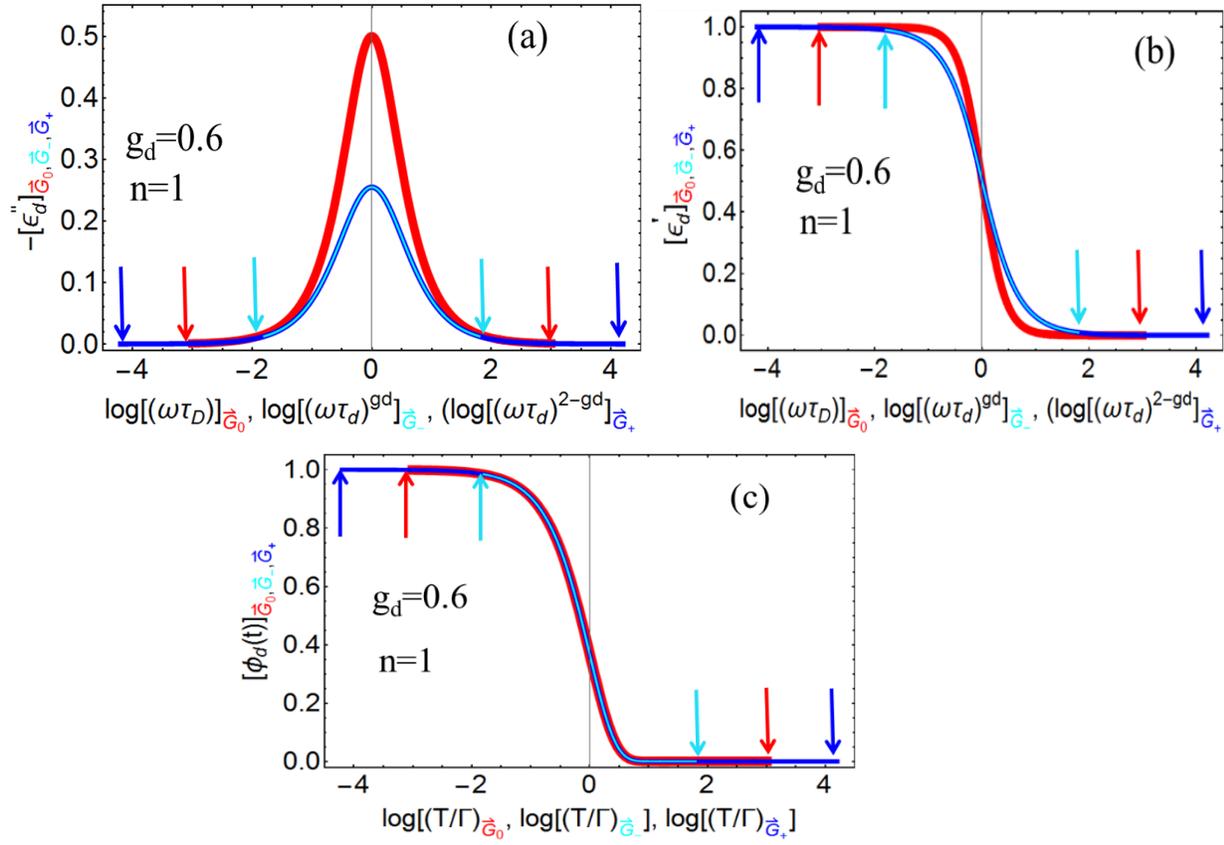

FIG. 2. (a) Shows the normalized log-linear plots of dielectric loss, (b) dielectric constant based on Eqs. (3), (4) and (5) and (c) the relaxation functions based on Eqs. (8), (9) and (10) for $g_d=1$ (Debye), and $g_d=0.6$.

Three normalized log-linear plots in (a) & (b) are:

(i) Thick red line in (a) and (b) are $-Im/Re[1/(1+i\omega\tau_D)]$ as a function of $\omega\tau_D$ based on Eq. (3) for the Debye dipole $\mathbf{G}_0$ process, where $g_d=1$.

(ii) The cyan line (on the blue line) in (a) and (b) are $-Im/Re[1/(1+i^{g_d}(\omega\tau_d)^{g_d})]$ as a function of $(\omega\tau_d)^{g_d}$ based on Eq. (4) for the non-Debye dipole $\mathbf{G}_-$ process, where x-axis variable $(\omega\tau_d)^{g_d}$ is compressed when $0<g_d<1$ with respect to $(\omega\tau_d)$. As $g_d$ decreases, the compression is enhanced with respect to the Debye dielectric loss process.

(iii) The blue line in (a) and (b) are $-Im/Re[1/(1+i^{g_d}(\omega\tau_d)^{2-g_d})]$ as a function of $(\omega\tau_d)^{2-g_d}$ based on Eq. (5) for the non-Debye dipole $\mathbf{G}_+$, where x-axis variable $(\omega\tau_d)^{2-g_d}$ is stretched with respect to $(\omega\tau_d)$, where the exponent $(2-g_d)$ is $1<(2-g_d)<2$. As $g_d$ decreases, the stretching is enhanced with respect to the Debye dielectric process. The arrows indicate extent of compression and stretching are equal in magnitudes and the depression in dielectric loss and the shift in dielectric constant is depends on phase $i^{g_d}$. The compression and stretching are equal in magnitudes and it is due to redistribution and conservation of Debye dipole dielectric loss energy in terms dual dipole dielectric loss processes.

(c) Shows three normalized time domain log-linear plots based on Eqs. (8)-(10) for $g_d=1$, $g_d=0.6$. Three normalized plots in Fig. 2(c) are:

(i) The thick red line in (c) is $\exp[-(t/\tau_D)]$ as a function of $(t/\tau_D)$ based on Eq. (8) for the Debye dipole $\mathbf{G}_0$ process, where $g_d=1$.

(ii) The solid cyan line (on blue line) in (c) is $\exp[-(T/\Gamma)_{\mathbf{G}_-}]$ as a function of $(T/\Gamma)_{\mathbf{G}_-}$ based on Eq. (9) for the non-Debye dipole $\mathbf{G}_-$ process, where x-axis variable $(T/\Gamma)_{\mathbf{G}_-}$ is compressed by $0 < g_d < 1$. As $g_d$ decreases, the compression is enhanced.

(iii) The blue line in (c) is $\exp[-(T/\Gamma)_{\mathbf{G}_+}]$ as a function of $(T/\Gamma)_{\mathbf{G}_+}$ based on Eq. (10) for the non-Debye dipole $\mathbf{G}_+$ process, where x-axis variable $(T/\Gamma)_{\mathbf{G}_+}$ is stretched by $0 < g_d < 1$. As $g_d$ decreases, stretching is enhanced. The arrows indicate extent of compression and stretching and these are equal in magnitudes. In both time and frequency domains the compression and stretching are equal in magnitudes and it is due to redistribution and conservation Debye dipole dielectric loss energy. As in case of Debye dielectric loss, dual dipole dielectric loss is also frequency dependent and the dual dipole dielectric loss peak happens to be at $\omega=1/\tau_d$, where $\tau_D=\tau_d$.

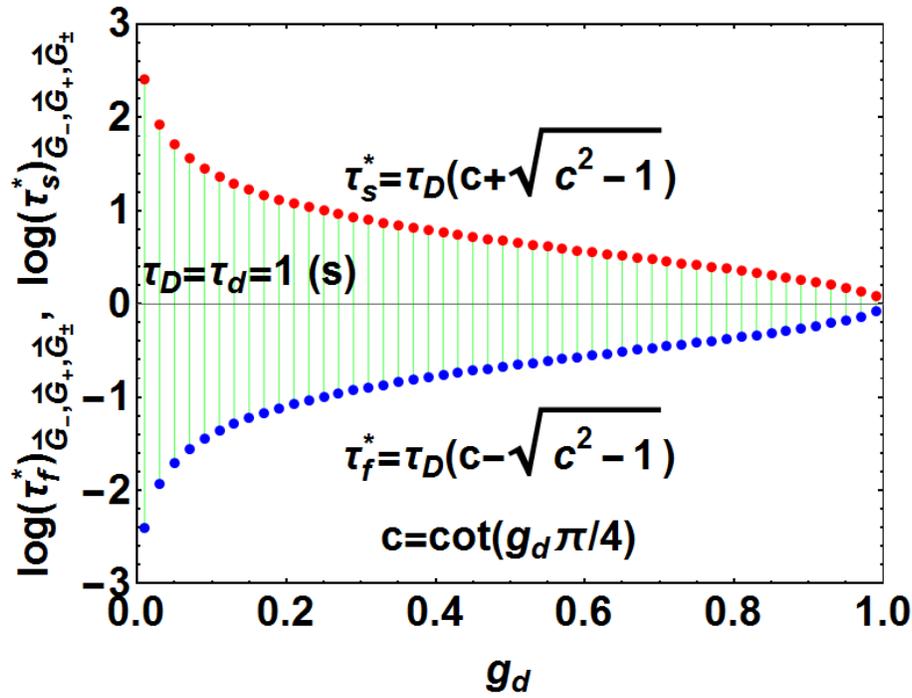

FIG. 3: Showing the slow relaxation time $\tau_s^*$, based on Eq. (6), and the fast relaxation time $\tau_f^*$, based on Eq.(7), as a function of interaction strength $g_d$ for the non-Debye dipole $\mathbf{G}_-$ or $\mathbf{G}_+$ or $\mathbf{G}_-+\mathbf{G}_+$ process for $\tau_D=\tau_d=1(s)$. The $\tau_s^*$, and $\tau_f^*$, are left and right sides of the Debye dielectric loss curve and $\omega_s^*=1/\tau_s^*$, and $\omega_f^*=1/\tau_f^*$, are right and left sides of the Debye dielectric loss curve. The shift in the relaxation time on both sides with respect Debye relaxation $\tau_D$ is symmetrical and shows hysteresis structure with $\tau_s^*$ and $\tau_f^*$ in the real part of complex dielectric and in the polarization. The magnitude of shift depends on strength of non-Debye dipole $g_d$. This is the novel and startling result in the proposed model.

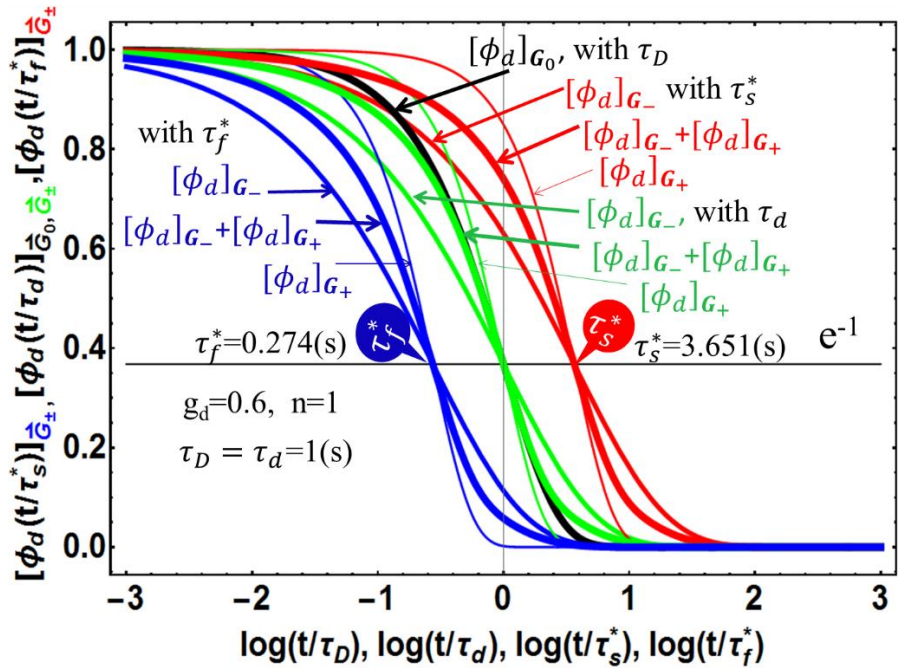

FIG. 4. Black thick line is the relaxation function $\exp[-(t/\tau_D)]$, (Debye, $g_d=1$) based on Eq. (8). In the middle, green, thick, thin and thicker lines are relaxation functions (i) $\exp[-(t/\tau_d)^{g_d}]$ based on Eq. (9) (ii) $\exp[-(t/\tau_d)^{2-g_d}]$ based on Eq. (10) and (iii) $(1/2)(\exp[-(t/\tau_d)^{g_d}]+\exp[-(t/\tau_d)^{2-g_d}])$ based on Eq. (13), for n=1, $\tau_D=\tau_d=1$(s), as a function of $(t/\tau_d)$ for the dipoles $\mathbf{G}_-$, $\mathbf{G}_+$ and $\mathbf{G}_-+\mathbf{G}_+$ processes respectively. On the left, red, thick, thin and thicker lines are relaxation functions (i) $\exp[-(t/\tau_s^*)^{g_d}]$, (ii) $\exp[-(t/\tau_s^*)^{2-g_d}]$, based on Eq. (11) and (iii) $(1/2)(\exp[-(t/\tau_s^*)^{g_d}]+\exp[-(t/\tau_s^*)^{2-g_d}])$, based on Eq. (13), for n=1, $\tau_D=\tau_d=1$(s), as a function of $(t/\tau_s^*)$ for the dipoles $\mathbf{G}_-$, $\mathbf{G}_+$ and $\mathbf{G}_-+\mathbf{G}_+$ processes respectively with non-Debye slow relaxation time $\tau_s^*$. On the right, blue, thick, thin and thicker lines are relaxation function (i) $\exp[-(t/\tau_f^*)^{g_d}]$, (ii) $\exp[-(t/\tau_f^*)^{2-g_d}]$, based on Eq. (11) and (iii) $(1/2)(\exp[-(t/\tau_f^*)^{g_d}]+\exp[-(t/\tau_f^*)^{2-g_d}])$, based on Eq. (13), for n=1, $\tau_D=\tau_d=1$(s), as a function of $(t/\tau_f^*)$ for the dipoles $\mathbf{G}_-$, $\mathbf{G}_+$ and $\mathbf{G}_-+\mathbf{G}_+$ processes respectively with fast non-Debye relaxation time $\tau_f^*$. The location of $\tau_s^*$ and $\tau_f^*$ are indicated for $g_d$=0.6 and they move away from $\tau_D$ on both sides as a function $g_d$. All three colored lines merge to Debye black line as $g_d$ becomes 1.

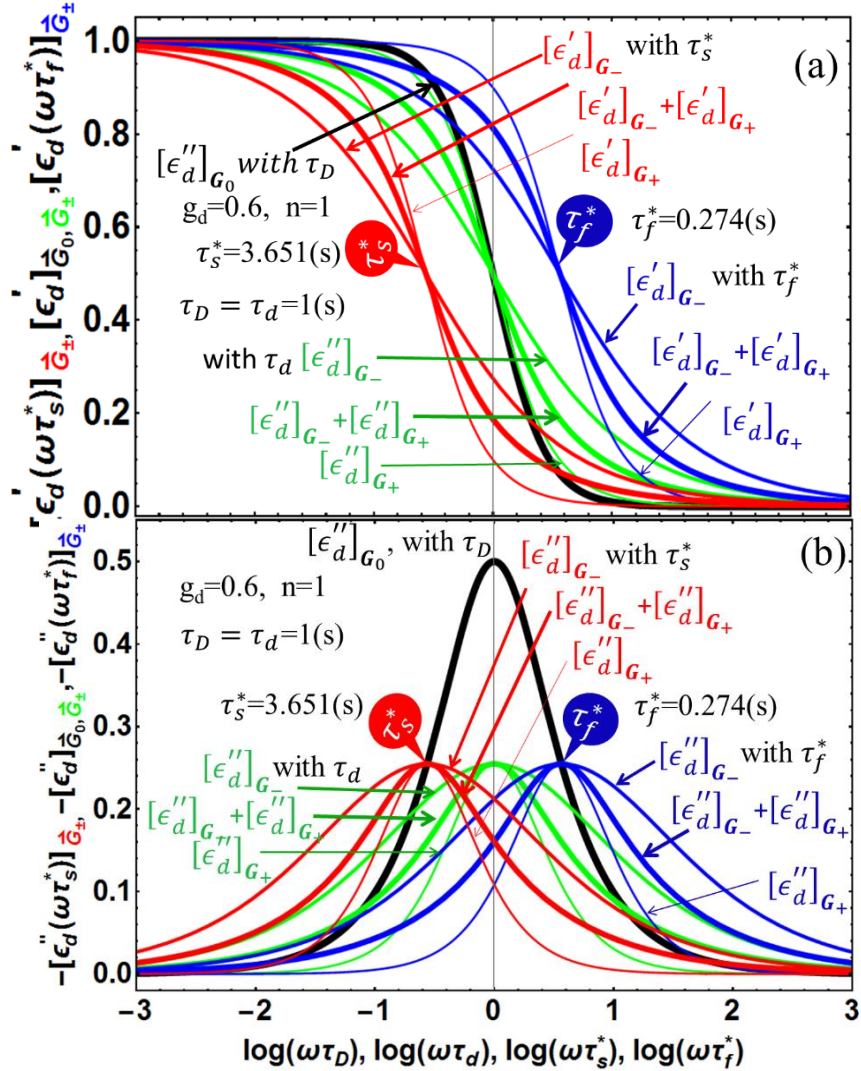

FIG. 5. (a) Shows the normalized functions of dielectric constant and (b) dielectric loss. Black line is dielectric constant and dielectric loss plot of $1/(1+i\omega\tau_D)$ (Debye, $g_d=1$), based on Eq. (3). In (a) and (b), the middle, green, thick, thin and thicker lines are dielectric constant and dielectric loss plot of (i) $1/(1+(i\omega\tau_d)^{g_d})$, based on Eq. (11) (ii) $1/(1+i^{g_d}(\omega\tau_d)^{2-g_d})$, based on Eq. (11) and (iii) $(1/2)(1/(1+(i\omega\tau_d)^{g_d})+1/(1+i^{g_d}(\omega\tau_d)^{2-g_d})$, based on Eq. (12), for n=1, $\tau_D=\tau_d=1(s)$, as a function of $(t/\tau_d)$ for the dipoles $\mathbf{G_-}$, $\mathbf{G_+}$ and $\mathbf{G_-}+\mathbf{G_+}$ processes respectively. On the left side, red thick, thin and thicker lines are dielectric constant and dielectric loss plot of (i) $1/(1+(i\omega\tau_s^*)^{g_d})$, (ii) $1/(1+i^{g_d}(\omega\tau_s^*)^{2-g_d})$, based on Eq. (11) and (iii) $(1/2)(1/(1+(i\omega\tau_s^*)^{g_d})+1/(1+i^{g_d}(\omega\tau_s^*)^{2-g_d})$, based on Eq. (13), for n=1, $\tau_D=\tau_d=1(s)$, as a function of $(t/\tau_s^*)$ for the dipole $\mathbf{G_-}$, $\mathbf{G_+}$ and $\mathbf{G_-}+\mathbf{G_+}$ processes respectively. On the right side, blue, thick, thin and thicker lines are dielectric constant and dielectric loss plots of (i) $1/(1+(i\omega\tau_f^*)^{g_d})$, based on Eq. (11) (ii) $1/(1+i^{g_d}(\omega\tau_f^*)^{2-g_d})$, based on Eq. (11) (iii) $(1/2)(1/(1+(i\omega\tau_f^*)^{g_d})+1/(1+i^{g_d}(\omega\tau_f^*)^{2-g_d})$, based on Eq. (13), for n=1, $\tau_D=\tau_d=1(s)$, as a function of $(t/\tau_f^*)$ for the dipoles $\mathbf{G_-}$, $\mathbf{G_+}$ and $\mathbf{G_-}+\mathbf{G_+}$ processes respectively. All three colored lines merge to Debye black line as $g_d$ becomes 1.

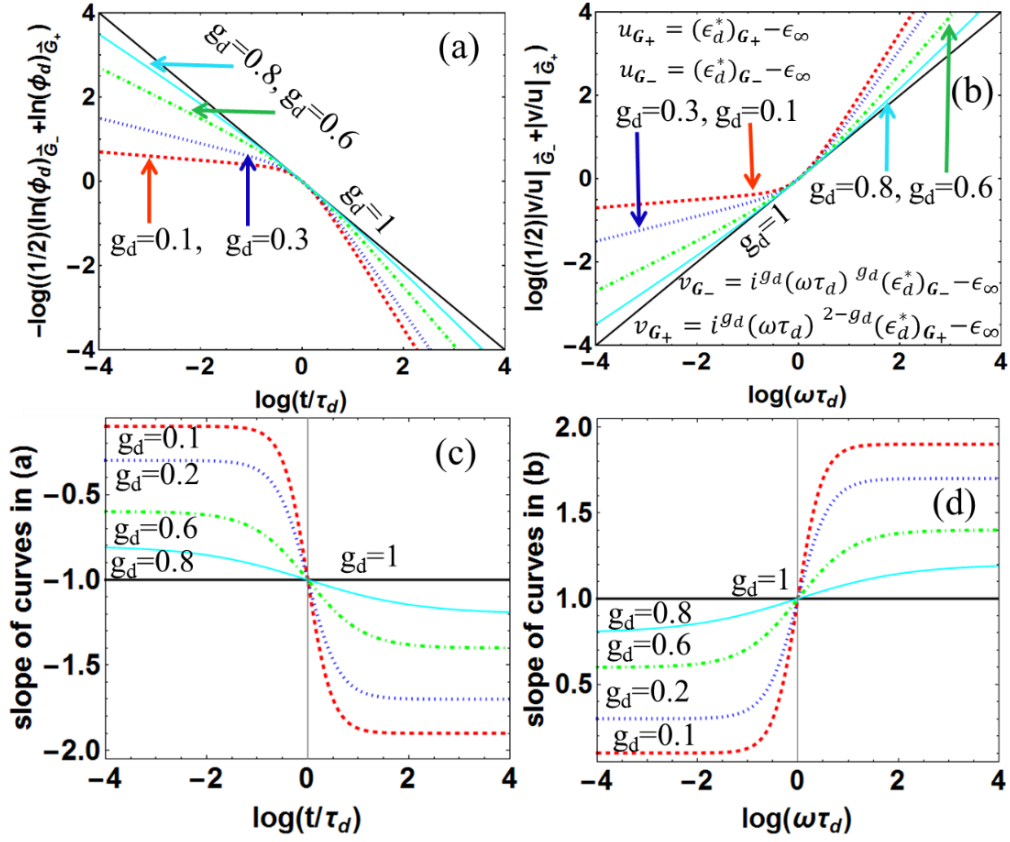

FIG. 6. (a) Shows the normalized log-log plot relaxation function and (b) complex dielectric function and (c) and (d) are their log-log slopes. The dielectric function, $\epsilon^*_{GG}(\omega)$, based on Eq. (12) and the relaxation function $\phi_{GG}(\omega)$, based on Eq. (13) can be written as:

$\ln(\exp[-(t/\tau_d)^{g_d}]) + \ln(\exp[-(t/\tau_d)^{2-g_d}]) = -((t/\tau_d)^{g_d} + (t/\tau_d)^{2-g_d})$.

$|v/u|_{G_-} + |v/u|_{G_+} = (\omega\tau_d)^{g_d} + (\omega\tau_d)^{2-g_d}$,

where

$v_{G_-} = i^{g_d}(\omega\tau_d)^{g_d}(\epsilon^*_d)_{G_-} - \epsilon_\infty$ , $u_{G_-} = (\epsilon^*_d)_{G_-} - \epsilon_\infty$,

$v_{G_+} = i^{g_d}(\omega\tau_d)^{2-g_d}(\epsilon^*_d)_{G_+} - \epsilon_\infty$ , $u_{G_+} = (\epsilon^*_d)_{G_+} - \epsilon_\infty$

The log-log plot of relaxation function, dielectric function and their slopes are shown in (a)-(d).

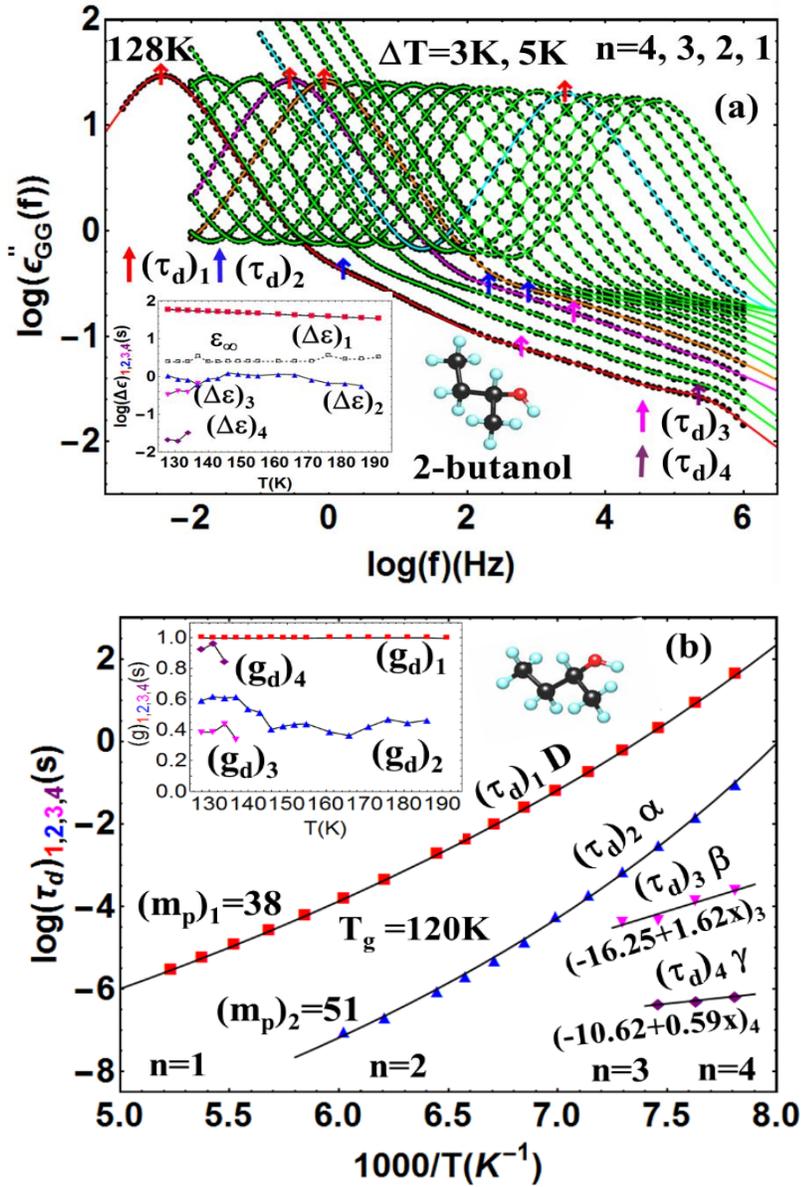

FIG. 7. (a) Color lines are fit results of dielectric loss based on Eq. (12) for 2-butanol with n=4 (red onwards), 3 (magenta), 2 (orange onwards), 1 (cyan) for T=128-191K, $\Delta T$=3K up to 161 and $\Delta T$=5K for T>161K. Arrows indicate relaxation time: n=1, Debye $(\tau_d)_1$; n=2, $\alpha$, $(\tau_d)_2$; n=3, $\beta$ $(\tau_d)_3$, n=4, $\gamma$, $(\tau_d)_4$ of each subunit motion. The inset shows Debye dielectric strength $(\Delta\varepsilon)_1$-$(\Delta\varepsilon)_4$ (in log scale) of subunits, their sum and $\varepsilon_\infty$. (b) Shows the temperature dependence $(\tau_d)_1$-$(\tau_d)_4$ obtained based on Eq. (12). The VFT, fragility based on VFT function, Arrhenius fit parameters are indicated for $(\tau_d)_1$-$(\tau_d)_4$, where n=1, Debye; n=2, $\alpha$; n=3, $\beta$, n=4, $\gamma$. The inset shows the interaction strength $(g_d)_1$-$(g_d)_4$ of subunits as function of temperature. The excess wing is observed in $\beta$ (n=3) process due to the strong interaction strength of $(g_d)_3\sim0.4$ for initial three temperatures. The schematic molecular structure is shown in the figures. In the sketch of the molecular structure, the oxygen atom is highlighted in red.

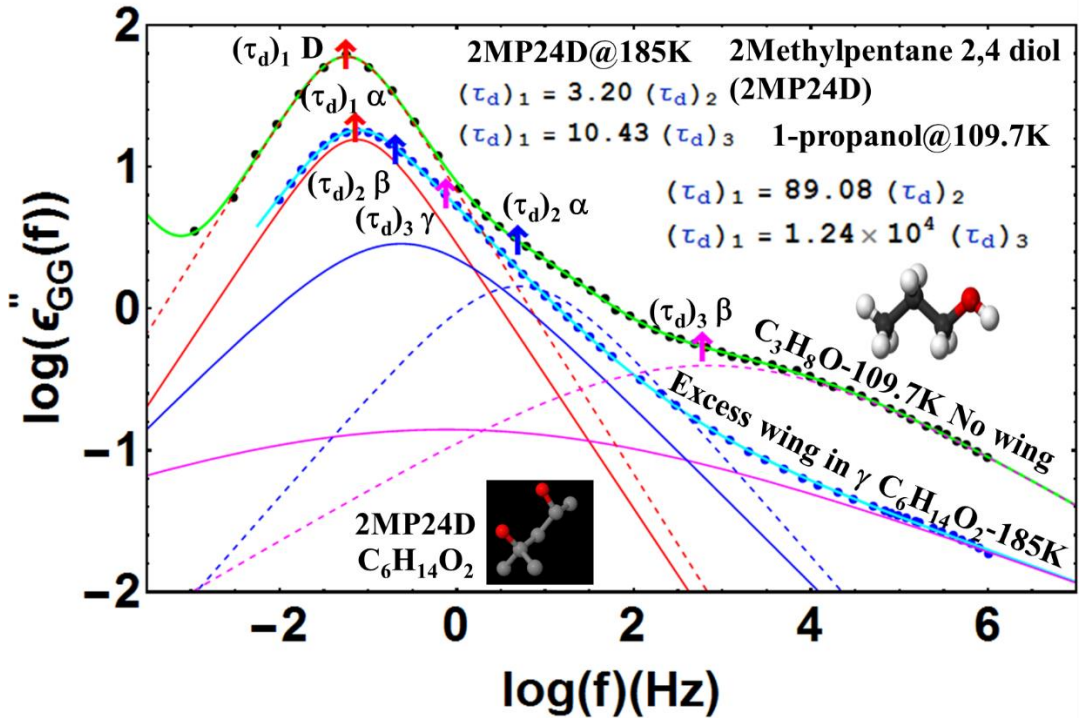

FIG. 8. Fit results for the dielectric loss based on Eq. (12) for n=3, for the 2 methyl pentane 2,4 diol (2MP24D) for T=185K and 1-propanal for 109.7K, where n=1, Debye (for 1-propanol) $(\tau_d)_1$; n=2, $\alpha$, $(\tau_d)_2$; n=3, $\beta$, $(\tau_d)_3$. In both systems individual contributions of $(G_-)_{1,2,3}$ are shown as red, blue and magenta lines for 2MP24D, dashed lines for 1-propanol. The arrows on the dielectric loss curves indicate the relaxation time $(\tau_d)_{1,2,3}$. The relationship between $\alpha$, $\beta$ and $\gamma$ in 2MP24D indicate that these are closely spaced each other, whereas in the case of 1-propanol, these are well separated as indicated in figure. The excess wing is seen clearly in 2MP24D in $\gamma$ process, since interaction strength $(g_d)_3$ ~0.2, whereas in 1-propanol, interaction strength is ~0.4 for $\beta$ process with well separated dielectric loss peak and hence no excess wing. The schematic molecular structure is shown in the figures. In the sketch of the molecular structure, the oxygen atom is highlighted in red.

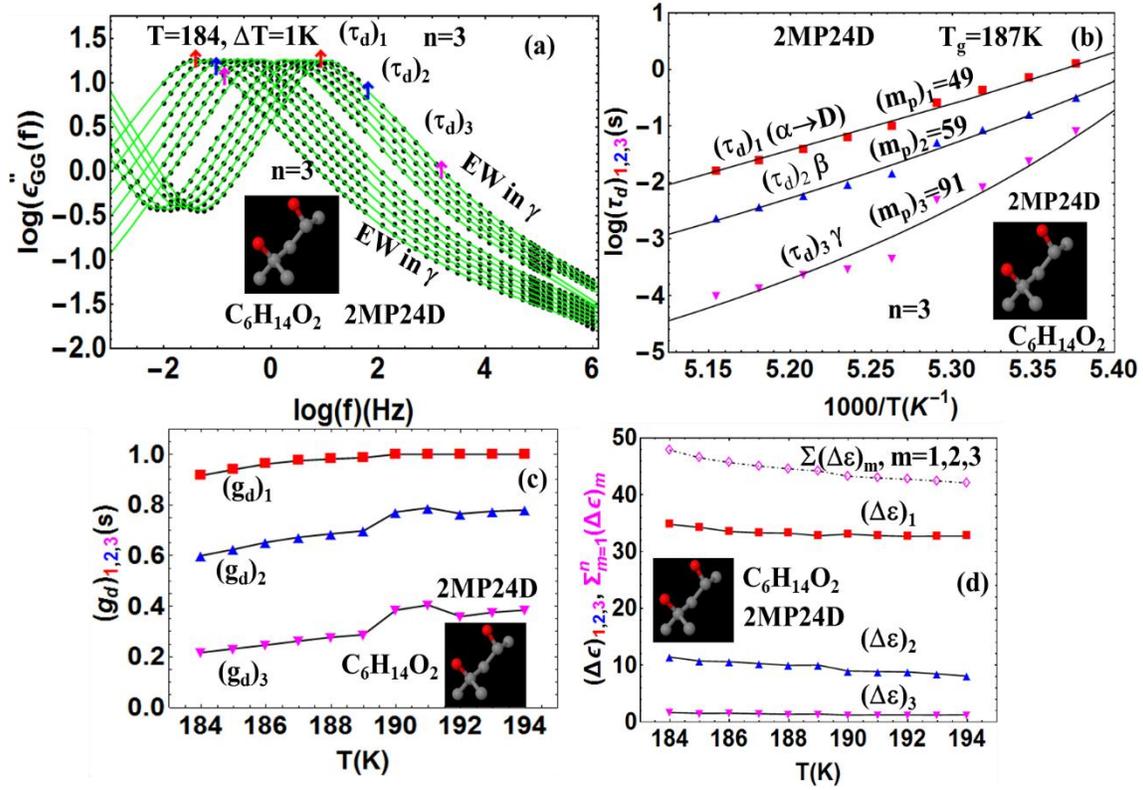

FIG. 9. (a) Green lines are fit results of dielectric loss based on Eq. (12) for 2 methyl pentane 2,4 diol (2MP24D) with n=3 for T=184-194K, $\Delta$T=1K. Arrows indicate relaxation time $(\tau_d)_1$-$(\tau_d)_3$: n=1, $\alpha \rightarrow$Debye, $(\tau_d)_1$; n=2, $\beta$, $(\tau_d)_2$; n=3, $\gamma$, $(\tau_d)_3$ of each subunit motion. (b) Shows the temperature dependence $(\tau_d)_1$-$(\tau_d)_3$ obtained based on Eq. (12). The VFT, fragility based VFT function, fit parameters are indicated for $(\tau_d)_1$-$(\tau_d)_3$, (c) Shows the interaction strength $(g_d)_1$-$(g_d)_3$ of subunits as function of temperature. The excess wing is observed in $\gamma$ (n=3) process due to the strong interaction strength of $(g_d)_3 \sim$0.2-0.4 for different temperatures. (d) Debye dielectric strength $(\Delta\varepsilon)_1$-$(\Delta\varepsilon)_3$ of subunits, their sum are shown as a function of temperatures. The schematic molecular structure is shown in the figures. In the sketch of the molecular structure, the oxygen atom is highlighted in red.

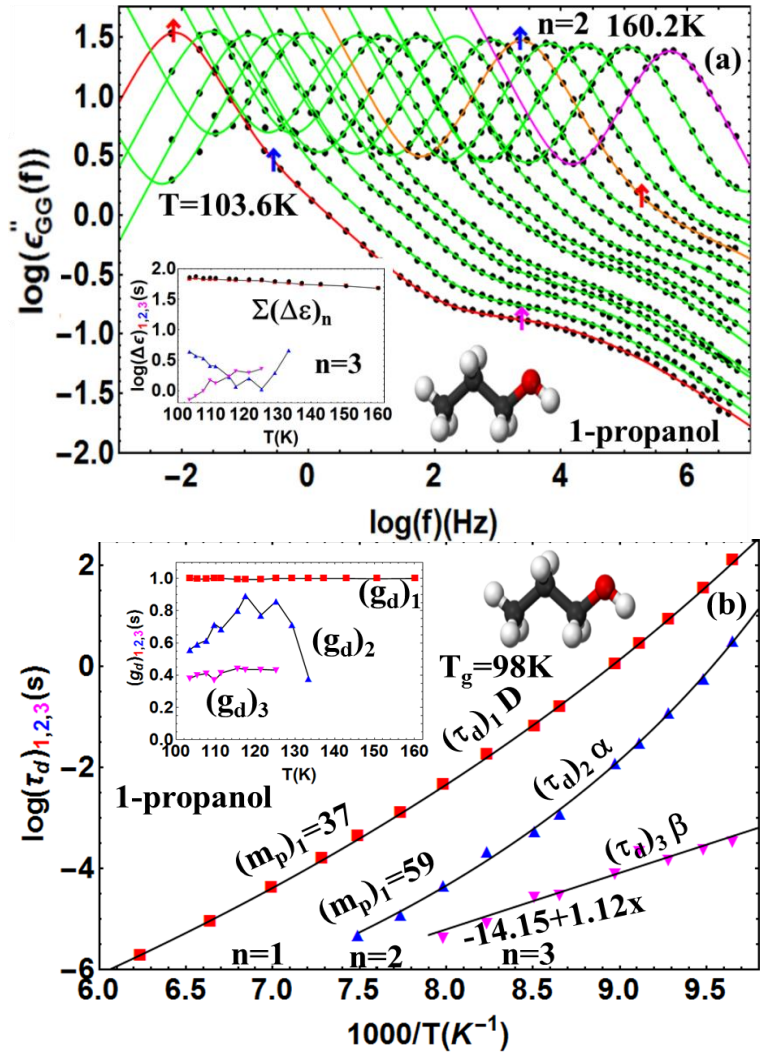

FIG. 10. (a) The color lines are fit results of dielectric loss based on Eq. (12) for 1-propanol with n=3 for T=103.6-160.2K. Arrows indicate relaxation time: n=1, Debye, $(\tau_d)_1$; n=2, $\alpha$, $(\tau_d)_2$, n=3, $\beta$, $(\tau_d)_3$ of each subunit motion. (b) Shows temperature dependence $(\tau_d)_1$-$(\tau_d)_3$ obtained based on Eq. (12). The VFT, fragility based VFT function and Arrhenius fit parameters are indicated for $(\tau_d)_1$-$(\tau_d)_3$, where n=1, Debye; n=2, $\beta$; n=3, $\gamma$. The inset in (a) shows Debye dielectric strength $(\Delta\varepsilon)_1$-$(\Delta\varepsilon)_3$ of subunits, their sum as a function of temperatures. The inset in (b) shows the interaction strength $(g_d)_1$-$(g_d)_3$ of subunits as function of temperature. The excess wing with shoulder is observed in $\beta$ ($(\tau_d)_3$, n=3) having strong interaction strength of $(g_d)_3$~0.4 for different temperatures. The schematic molecular structure is shown in the figures. In the sketch of the molecular structure, the oxygen atom is highlighted in red.

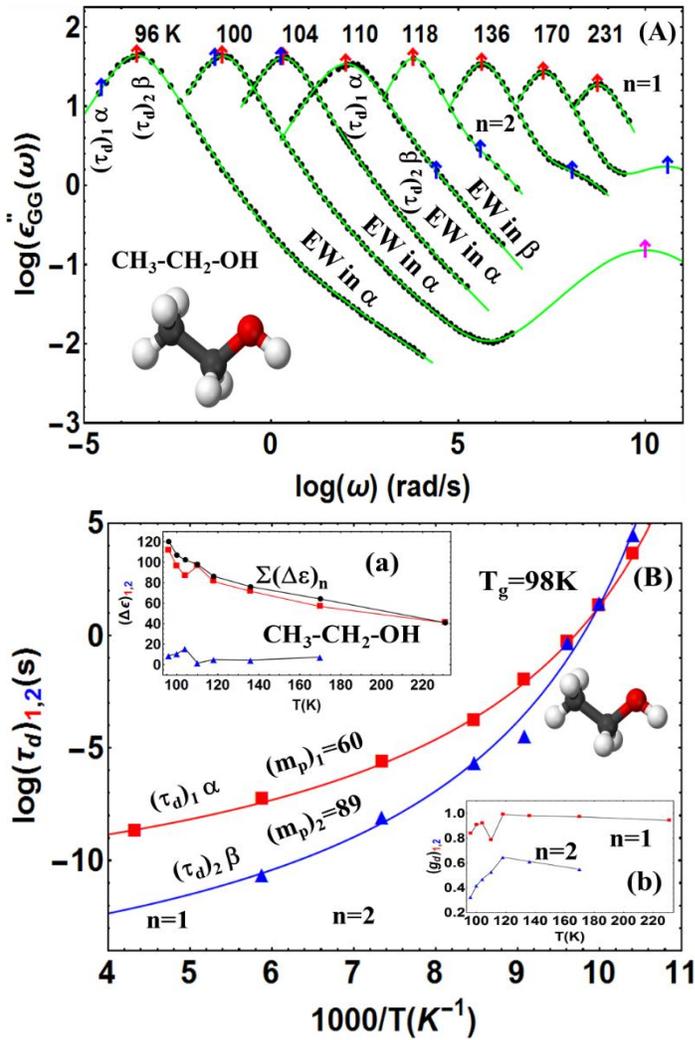

FIG. 11. (A) Green lines are fit results of dielectric loss based on Eq. (12) for ethanol with n=2 for T=96-226K. Arrows indicate relaxation time: n=1, α→Debye, $(\tau_d)_1$; n=2, β, $(\tau_d)_2$ of each subunit motion. (B) Shows the temperature dependence $(\tau_d)_1$-$(\tau_d)_2$ obtained based on Eq. (12). The VFT, fragility based VFT function are indicated for $(\tau_d)_1$-$(\tau_d)_2$, where n=1, α→Debye; n=2, β. The inset (a) in (B) shows Debye dielectric strength $(\Delta\varepsilon)_1$-$(\Delta\varepsilon)_2$ of subunits, their sum as a function of temperatures. The inset (b) in (B) shows the interaction strength $(g_d)_1$-$(g_d)_3$ of subunits as function of temperature. The excess wing is observed in α and as well as in β process due to the strong interaction strength of $(g_d)_3 \sim 0.2$-0.6 for different low temperatures. The schematic molecular structure is shown in the figures. In the sketch of the molecular structure, the oxygen atom is highlighted in red.

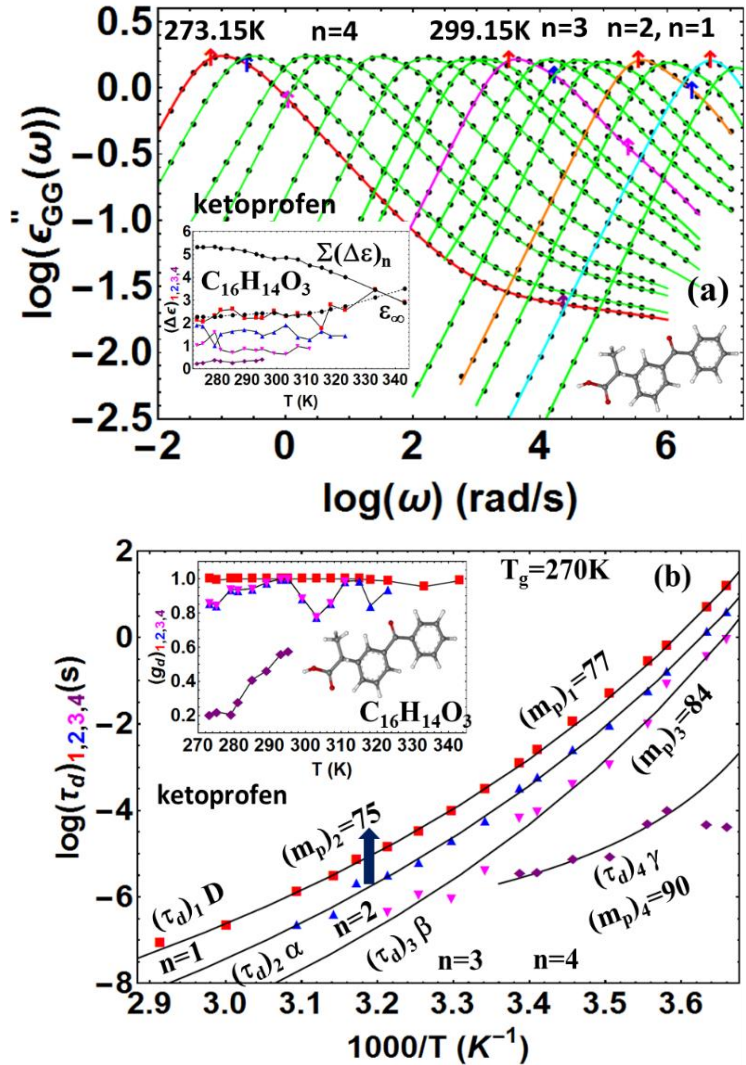

FIG. 12. (a) Color lines are fit results of dielectric loss based on Eq. (12) for ketoprofen drug with n=4 (red onwards), 3 (magenta), 2 (orange onwards), 1 (cyan) for T=273.15-343.15K. Arrows indicate relaxation time: n=1, Debye, $(\tau_d)_1$; n=2, $\alpha$, $(\tau_d)_2$; n=3, $\beta$, $(\tau_d)_3$; n=4, $\gamma$, $(\tau_d)_4$ of each subunit motion. The inset shows Debye dielectric strength $(\Delta\varepsilon)_1$-$(\Delta\varepsilon)_4$ of subunits, their sum and $\varepsilon_\infty$. (b) Show the temperature dependence $(\tau_d)_1$-$(\tau_d)_4$ obtained based on Eq. (12). The VFT, fragility based VFT function are indicated for $(\tau_d)_1$-$(\tau_d)_4$, where n=1, Debye; n=2, $\alpha$; n=3, $\beta$, n=4, $\gamma$. The inset shows the interaction strength $(g_d)_1$-$(g_d)_4$ of subunits as function of temperature. The excess wing is observed in $\gamma$ (n=4) process having strong interaction strength of $(g_d)_4\sim0.2$ for initial temperatures. The schematic molecular structure is shown in the figures. In the sketch of the molecular structure, the oxygen atom is highlighted in red.

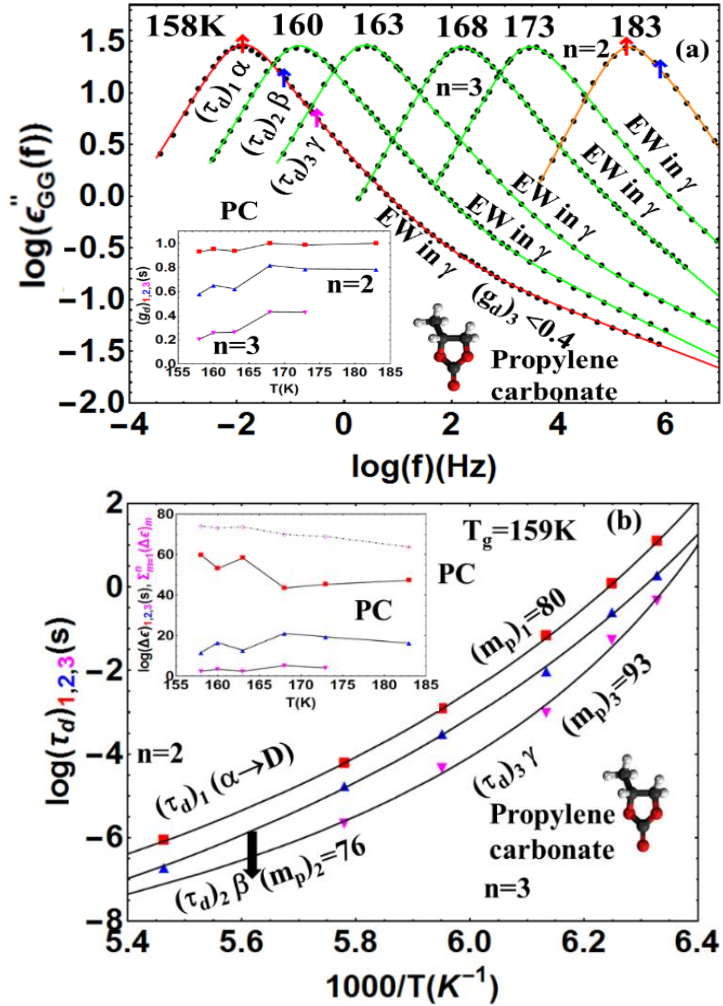

FIG. 13. (a) Color lines are fit results of dielectric loss based on Eq. (12) for propylene carbonate (PC) with n=3 (red onwards), 2 (orange), for T=158-183K. Arrows indicate relaxation time: n=1, α→Debye, $(\tau_d)_1$; n=2, β, $(\tau_d)_2$; n=3, γ, $(\tau_d)_3$ of each subunit motion. The inset shows the interaction strength $(g_d)_1$-$(g_d)_3$ of subunits as function of temperature. The excess wing is observed in γ (n=3) process having strong interaction strength of $(g_d)_3$~0.2-0.4 for initial five temperatures. (b) Shows the temperature dependence $(\tau_d)_1$-$(\tau_d)_3$ obtained based on Eq. (12). The VFT, fragility based VFT function fit parameters are indicated for $(\tau_d)_1$-$(\tau_d)_3$, where n=1, α→Debye; n=2, β; n=3, γ. The inset shows Debye dielectric strength $(\Delta\varepsilon)_1$-$(\Delta\varepsilon)_3$ of subunits, their sum. The schematic molecular structure is shown in the figures.

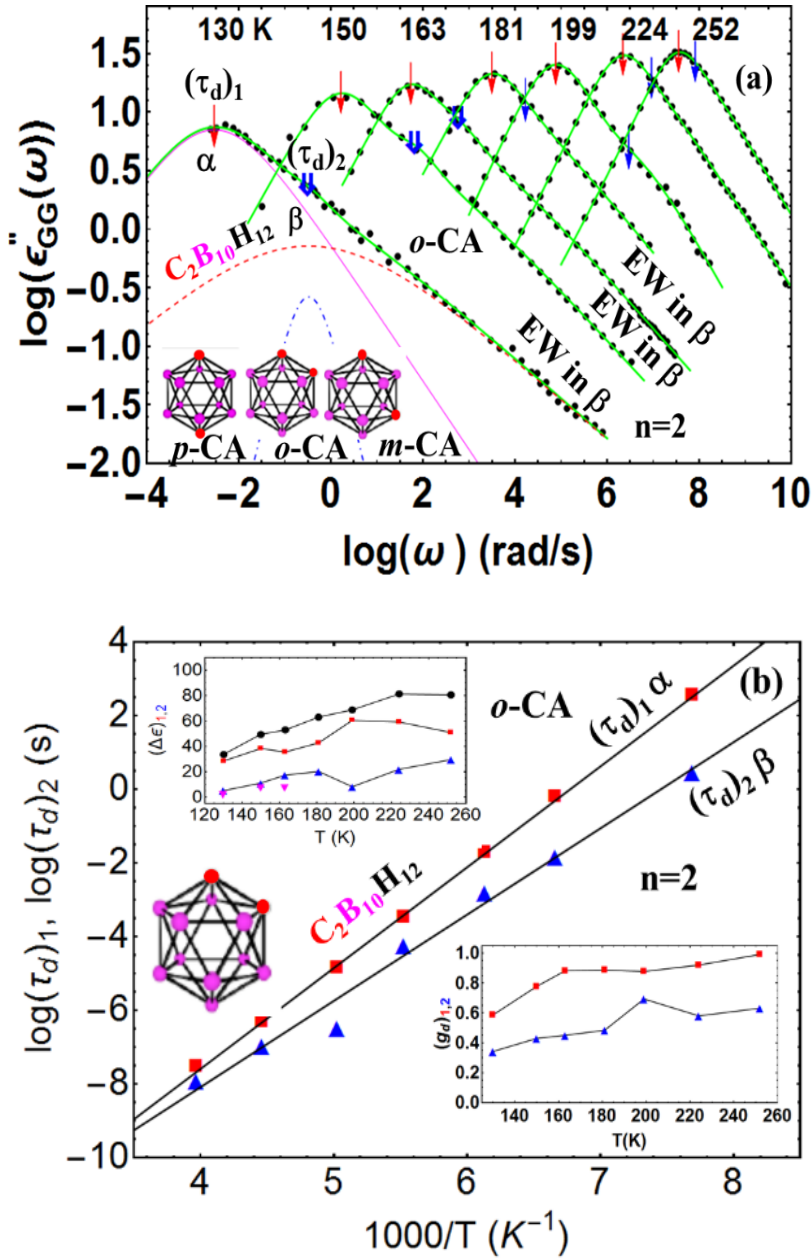

FIG. 14. (a) Green lines are fit results of dielectric loss based on Eq. (12) for carborane (*ortho*-CA) molecule with n=2 for T=130-252K. It forms an almost regularly shaped icosahedron corners are occupied by 10 boron and 2 carbon atoms and surrounded by 12 out-ward bonded hydrogen atom (not shown) (*ortho*-CA, 1 and 2 carbon position, *meta*-CA, 1 and 7 carbon position, *para*-CA, 1 and 12 carbon position with no net dipole moment). The schematic molecular structure is shown in the figures. Arrows indicate relaxation times: n=1, α, $(\tau_d)_2$; n=2, β, $(\tau_d)_2$ of each subunit motion. For T= 130, 150, 163K, β, $(\tau_d)_2$ has loss contributions from $(G_\pm)_2$ with excess wing (double arrow indicates for $(G_\pm)$). (b) is temperature dependence $(\tau_d)_1$-$(\tau_d)_2$ obtained based on Eq. (12) shows Arrhenius behavior for $(\tau_d)_1$ and $(\tau_d)_2$, where n=1, α; n=2, β. The top inset shows Debye dielectric strength $(\Delta\varepsilon)_1$-$(\Delta\varepsilon)_2$ of subunits, their sum as a function of temperatures. The interaction strength $(g_d)_1$-$(g_d)_3$ of subunits as function of temperature is shown as bottom inset in (b). There is excess wing in β process, since interaction strength is strong, where $(g_d)_2$~0.3 to 0.45 in low temperature and becomes 0.5 at high temperatures.

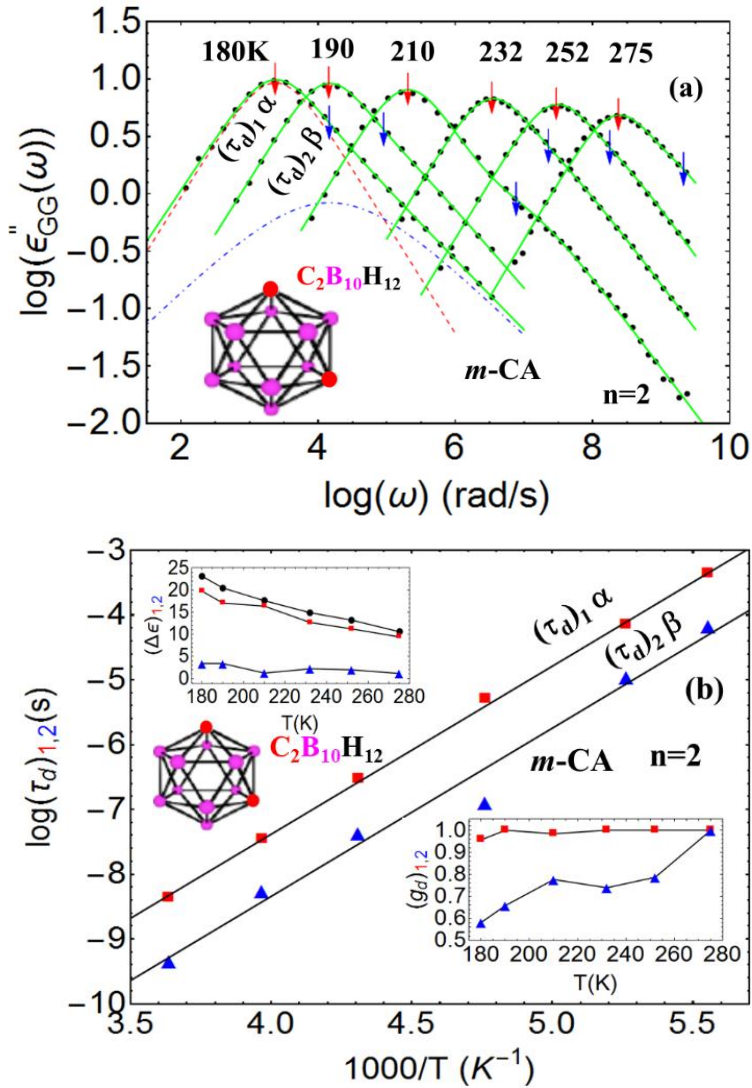

FIG. 15. (a) Green lines are fit results of dielectric loss based on Eq. (12) for meta-carborane (*m*-CA, 1 and 7 carbon position) with n=2 for T=180-275K. Arrows indicate relaxation times: n=1, α→Debye, $(\tau_d)_1$; n=2, β, $(\tau_d)_2$ of each subunit motion. (b) Show the temperature dependence $(\tau_d)_1$-$(\tau_d)_2$ obtained based on Eq. (12) shows Arrhenius fit for $(\tau_d)_1$-$(\tau_d)_2$, where n=1, α→Debye; n=2, β. The top inset shows Debye dielectric strength $(\Delta\varepsilon)_1$-$(\Delta\varepsilon)_2$ of subunits, their sum as a function of temperatures. The interaction strength $(g_d)_1$-$(g_d)_3$ of subunits as function of temperature is shown as bottom inset in (b). There is no excess wing in β process, since interaction strength is week, where $(g_d)_2$~0.6 in low temperature and becomes Debye at high temperatures. The schematic molecular structure is shown in the figures.

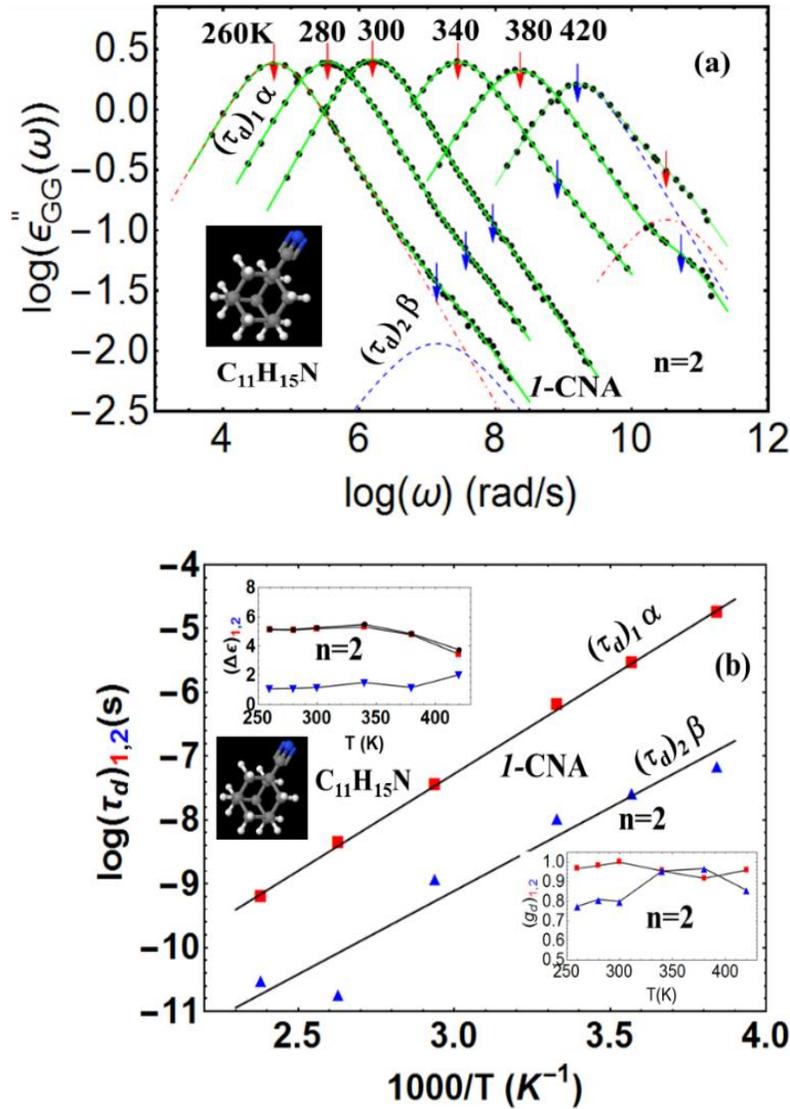

FIG. 16. (a) Green lines are fit results of dielectric loss based on Eq. (12) for 1-cyanoadamantane (*1*-CNA) with n=2 for T=260-420K. Arrows indicate relaxation times: n=1, α, $(\tau_d)_1$; n=2, β, $(\tau_d)_2$ of each subunit motion. (b) Shows the temperature dependence $(\tau_d)_1$-$(\tau_d)_2$ obtained based on Eq. (12) shows Arrhenius fit for $(\tau_d)_1$-$(\tau_d)_2$, where n=1, α; n=2, β. The top inset shows Debye dielectric strength $(\Delta\varepsilon)_1$-$(\Delta\varepsilon)_2$ of subunits, their sum as a function of temperatures. The interaction strength $(g_d)_1$-$(g_d)_2$ of subunits as function of temperature is shown as bottom inset in (b). There is no excess wing in β process, since interaction strength is week, where $(g_d)_2$~0.6 in low temperature and becomes Debye at high temperatures. The schematic molecular structure is shown in the figures.

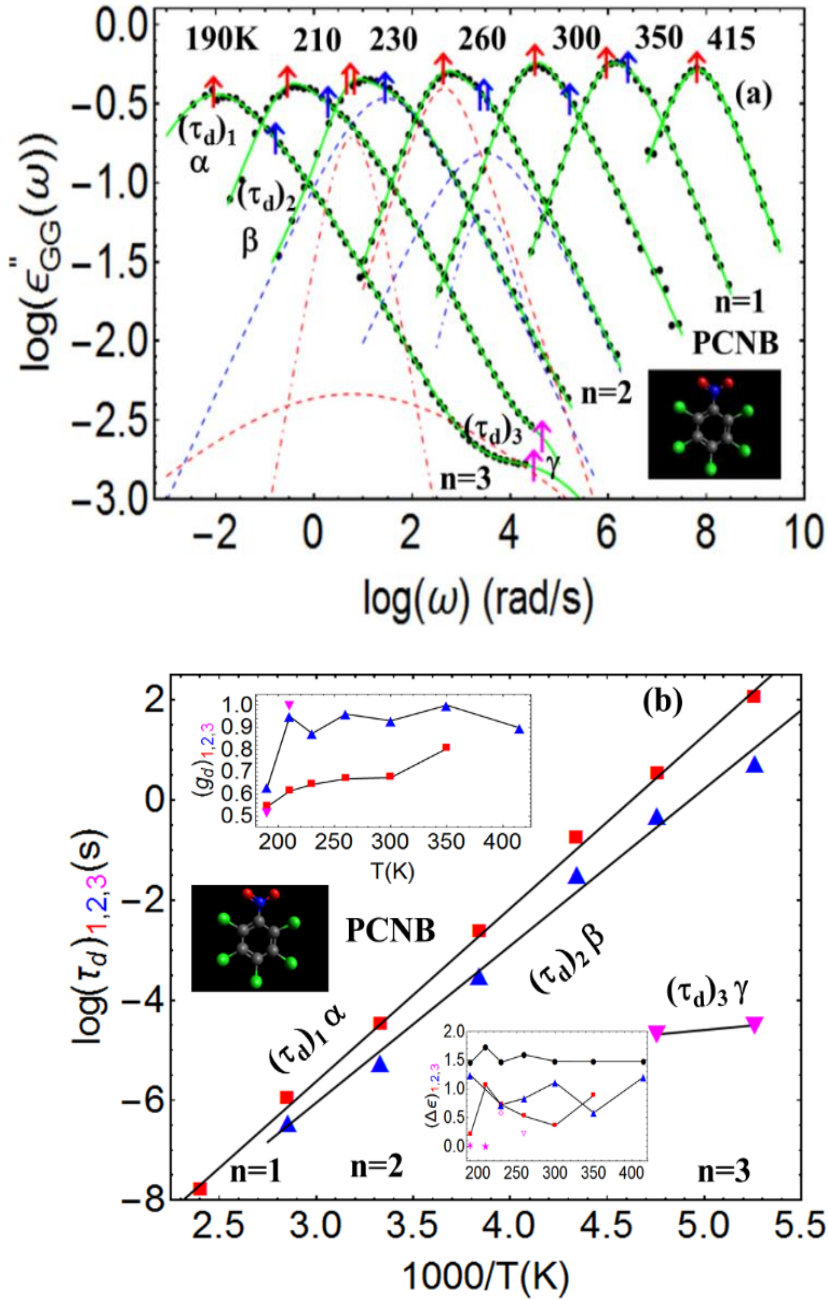

FIG. 17. (a) Green lines are fit results of dielectric loss based on Eq. (12) for pentachloronitrobenzene ($C_6Cl_5NO_2$) (PCNB) with n=3, 2 for T=190-415K. This molecule has circular disk structure. Initial temperatures T=190 and 210, three motions are observed with relaxation times $(\tau_d)_1$-$(\tau_d)_3$ as α, β and γ. For T=230K, α has loss contribution from $(G_\pm)_1$, and β has loss contribution from $(G_-)_1$ and at T=260K, β has loss contribution from $(G_\pm)$ and α has loss contribution from $(G_-)_1$. These individual loss contributions are shown as a dashed and a dot dashed lines. Arrows indicate relaxation times: n=1, α, $(\tau_d)_1$; n=2, β; $(\tau_d)_2$, n=3, γ, $(\tau_d)_3$ of each subunit motion (double arrow indicates for $(G_\pm)$). (b) Shows the temperature dependence $(\tau_d)_1$-$(\tau_d)_3$ obtained based on Eq. (12) and shows Arrhenius behavior for $(\tau_d)_1$-$(\tau_d)_2$, where n=1, α; n=2, β. The top inset shows Debye dielectric strength $(\Delta\epsilon)_1$-$(\Delta\epsilon)_2$ of subunits, their sum as a black dots as a function of temperatures. The interaction strength $(g_d)_1$-$(g_d)_2$ of subunits as function of temperature is shown as bottom inset in (b). There is no excess wing in β process since interaction strength is week, where $(g_d)_2$~0.6 in low temperature and becomes Debye at high temperatures. The schematic molecular structure is shown in the figures.

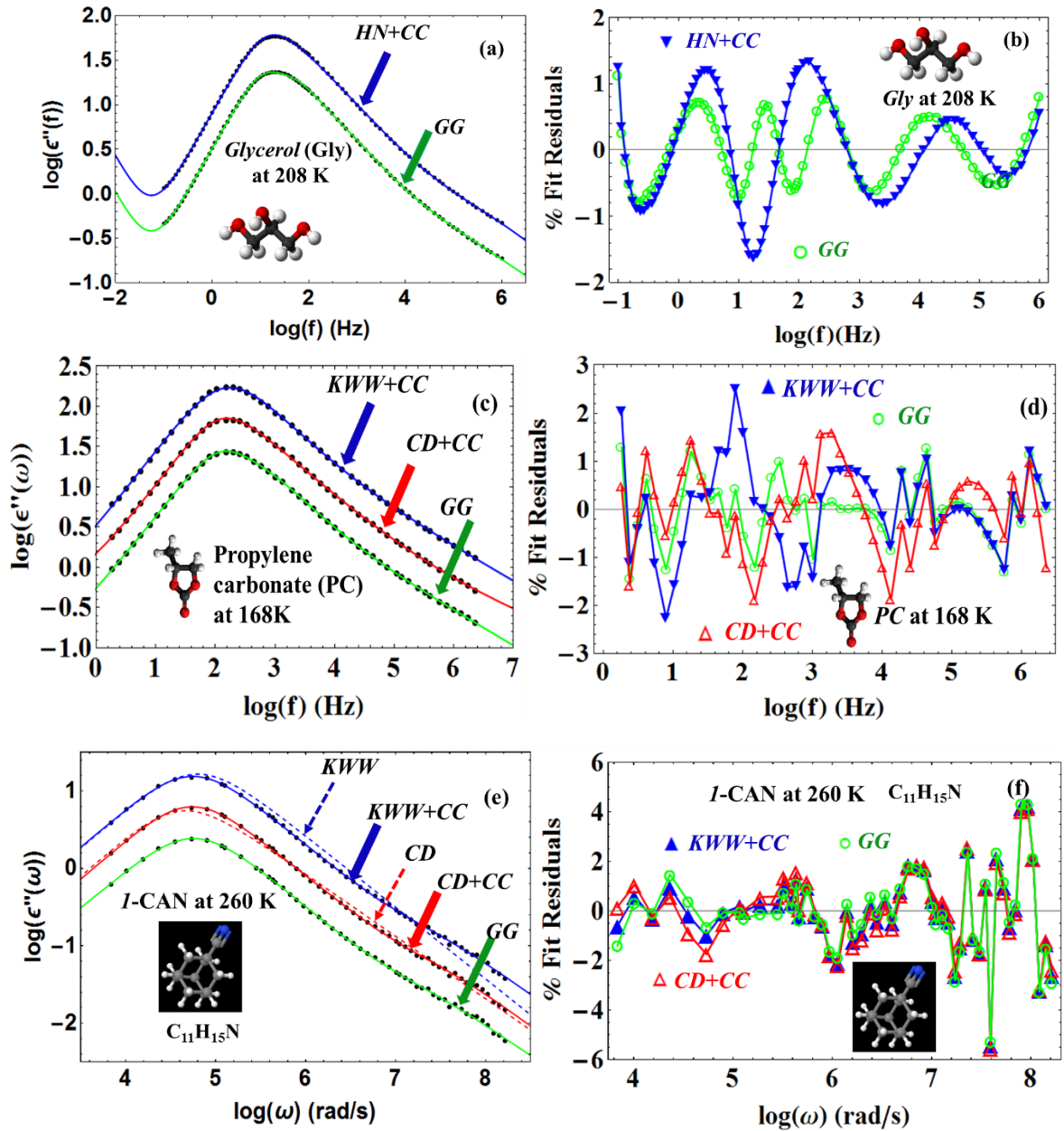

FIG. 18. (a) Shows the fit results (b) fit residuals of HN+CC (blue) and present proposed model GG (green, n=4) for glycerol (Gly) at 208K. (c) Shows the fit results and (b) fit residuals of KWW+CC (blue), CD+CC (red) and present proposed GG model (green, n=3) for propylene carbonate (PC) at 168K. (e) Shows fit results and (f) fit residues of KWW+CC (blue line), CD+CC (red line) and present proposed GG model (green, n=2) for 1-cynaoadmaentane (1-CAN) at 260K, where y scale is shifted by 0.4 units for each model starting from GG. The KWW and CD functions show very poor fitting with experimental values and hence fit residues are not shown. The least magnitude of fit residues is observed for the present proposed GG model.